\documentclass[pra,aps,amssymb,twocolumn]{revtex4}
\usepackage{graphicx}
\newcommand{\half}{{\frac{1}{2}}}
\newcommand{\ket}[1]{|#1\rangle}
\newcommand{\bra}[1]{\langle#1|}
\newcommand{\braket}[2]{\langle#1|#2\rangle}
\newcommand{\mr}{\mathrm}
\newcommand{\mc}{\mathcal}
\newcommand{\tr}{\mathrm{Tr}}

\begin{document}

\title{The physics of Maxwell's demon and information}
\author{Koji Maruyama$^{1,2}$, Franco Nori$^{1,2,3}$, and Vlatko Vedral$^{4,5}$}
 \affiliation{$^1$Advanced Science Institute, RIKEN (The Institute of Physical and Chemical
Research), Wako-shi 351-0198, Japan\\
$^2$CREST, Japan Science and Technology Agency (JST), Kawaguchi, Saitama 332-0012, Japan\\
$^3$Center for Theoretical Physics, Physics Department, Center for the Study of Complex Systems,
University of Michigan, Ann Arbor, MI 48109-1040, USA \\
$^4$The School of Physics and Astronomy, University of Leeds, Leeds, LS2 9JT, UK \\
$^5$Quantum Information Technology Lab, National University of Singapore, 117542 Singapore,
Singapore}
\begin{abstract} Maxwell's demon was born in 1867 and still thrives in modern physics. He plays
important roles in clarifying the connections between two theories: thermodynamics and
information. Here, we present the history of the demon and a variety of interesting consequences
of the second law of thermodynamics, mainly in quantum mechanics, but also in the theory of
gravity. We also highlight some of the recent work that explores the role of information,
illuminated by Maxwell's demon, in the arena of quantum information theory.
\end{abstract}
\date{\today}
\maketitle

\tableofcontents

\section{Introduction}\label{intro}

The main focus of this article is the second law of thermodynamics in terms of information. There
is a long history concerning the idea of information in physics, especially in thermodynamics,
because of the significant resemblance between the information theoretic \textit{Shannon entropy}
and the thermodynamic \textit{Boltzmann entropy}, despite the different underlying motivations and
origins of the two theories (See for example Ref.~\cite{leff03} and references therein). A
definitive discovery in this context is Landauer's \textit{erasure principle}, which clearly
asserts the relationship between information and physics. As we will see below, having been
extended to the quantum regime by Lubkin \cite{lubkin87}, it has proven useful in understanding
the constraints on various information processing tasks from a physical point of view.

Here we will review the background on the correspondence between information and physics, in
particular from the point of view of thermodynamics. Starting with the classic paradox of
Maxwell's demon, we will discuss the erasure principle from the perspectives that will be of
interest to us in later sections. Then we shall review a variety of consequences of the second
law, mainly in quantum mechanics, and also briefly in the theory of gravity. These are
thought-provoking because the second law of thermodynamics is a sort of meta-rule, which holds
regardless of the specific dynamics of the system we look at. Besides, the fundamental postulates
of quantum mechanics, as well as general relativity, do not presume the laws of thermodynamics in
the first place. After reviewing these classic works and appreciating how powerful and how
universal the second law is, we will discuss some of recent progress concerning the intriguing
interplay between information and thermodynamics from the viewpoint of quantum information theory.

\section{Maxwell's demon}
\subsection{The paradox}
A character who has played an important role in the history of physics, particularly in
thermodynamics and information, is Maxwell's demon. It was first introduced by Maxwell in 1871
\cite{maxwell867} to discuss the ``limitations of the second law of thermodynamics", which is also
the title of a section in his book. The second law (in Clausius's version) states \cite{pippard}:
\textit{``It is impossible to devise an engine which, working in a cycle, shall produce no effect
other than the transfer of heat from a colder to a hotter body."} Maxwell devised his demon in a
thought experiment to demonstrate that the second law is only a statistical principle that holds
\textit{almost} all the time, and not an absolute law set in stone.

The demon is usually described as an imaginary tiny being that operates a tiny door on a
partition which separates a box into two parts of equal volumes, the left and the right. The box
contains a gas which is initially in thermal equilibrium, i.e. its temperature $T$ is uniform over
the whole volume of the box. Let $\langle v\rangle_T$ denote the average speed of molecules that
form the gas. The demon observes the molecules in the left side of the box, and if he sees one
approaching the door with a speed less than $\langle v\rangle_T$, then he opens the door and lets
the molecule go into the right side of the box. He also observes the molecules in the right, and
if he sees one approaching with a speed greater than $\langle v\rangle_T$, then he opens the door
to let it move into the left side of the box.

Once he has induced a small difference in temperatures between the right and the left,
his action continues to transfer heat from the colder part (right) to the hotter part
(left) without exerting any work, thus he is breaking Clausius's form of the second law.
This type of demon is referred to as the \textit{temperature demon}.

There is another type of Maxwell's demon, who is ``less intelligent" than the temperature demon.
Such a demon merely allows all molecules moving in one direction to go through, while stopping all
those moving the other way to produce a difference in pressure. This \textit{pressure demon} runs
a cycle by making the gas interact with a heat bath at a constant temperature after generating a
pressure inequality. The sole net effect of this cycle is the conversion of heat transferred from
the heat bath to work. This is also a plain violation of the second law, which rules out perpetuum
mobile (in Kelvin's form): \textit{``It is impossible to devise an engine which, working in a
cycle, shall produce no effect other than the extraction of heat from a reservoir and the
performance of an equal amount of mechanical work."}

The second law can also be phrased as \textit{``in any cyclic process the total entropy of the
physical systems involved in the process will either increase or remain the same"}. Entropy is, in
thermodynamics, a state variable $S$ whose change is defined as $dS=\delta Q/T$ for a reversible
process at temperature $T$, where $\delta Q$ is the heat absorbed. Thus, irrespective of the type
of demon, temperature or pressure, what he attempts to do is to decrease the entropy of the whole
system for the cyclic process.

Historically, a number of physical mechanisms that might emulate the demon without any intelligent
beings have also been proposed. One notable example should be the trap-door model by Smoluchowski
\cite{smoluchowski12}. Instead of an intelligent demon operating the door, he considered a door
that is attached to the partition by a spring so that it only opens to one side, the left, say.
Then fast moving molecules in the right side can go into the left side by pushing the door, but
slow ones are simply reflected as the door is shut tightly enough for them and no molecules can go
into the right from the left. After a while, the temperature (as well as the density) of the left
side should become higher and the right side lower. Useful work would be extracted from this
spontaneously generated temperature difference. Smoluchowski pointed out that what prevents the
trap-door mechanism from achieving the demonic work are thermal fluctuations, i.e. Brownian
motion, of the door. The door might be kicked sometimes to let a fast molecule in; however, the
thermal fluctuations will lift the door up and let molecules go back to the right side, resulting
in no net temperature difference. This scenario was numerically analyzed in detail in Ref.
\cite{skordos92} to confirm the above reasoning. While there have been many other similar
mechanisms proposed, a more sophisticated model was discussed by Feynman as the ratchet-and-pawl
machine \cite{feynmanlecture}, which, again, does not work as a perpetual engine due to the
thermal fluctuations.

The demon puzzle, which had been a cardinal question in the theory of thermodynamics, is now why a
demon can never operate beyond the apparently fundamental limits imposed by the second law, no
matter how intelligent he is and no matter what type (temperature or pressure) he is. An ingenious
idea by Szilard treated the demon's intelligence as \textit{information} and linked it with
physics.

\subsection{Szilard's engine}\label{szilards_engine}
In 1929, the Hungarian scientist Leo Szilard presented a classical (non-quantum) analysis of
Maxwell's demon (pressure demon), formulating an idealized heat engine with one-molecule gas
\cite{szilard29}. Szilard's work was epoch-making in the sense that he explicitly pointed out, for
the first time, the significance of information in physics.

\begin{figure}
\begin{center}
\includegraphics[scale=0.35]{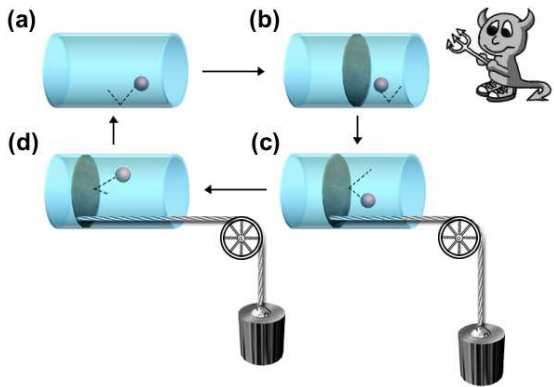}
\caption{Schematic diagram of Szilard's heat engine. A chamber of volume $V$ contains a
one-molecule gas, which can be found in either the right or the left part of the box. (a)
Initially, the position of the molecule is unknown. (b) Maxwell's demon inserts a partition at the
centre and observes the molecule to determine whether it is in the right or the left hand side of
the partition. He records this information in his memory. (c) Depending on the outcome of the
measurement (which is recorded in his memory), the demon connects a load to the partition. If the
molecule is in the right part as shown in the figure, he connects the load to the right hand side
of the partition. (d) The isothermal expansion of the gas does work upon the load, whose amount is
$kT\ln2$ which we call 1 bit. (Adapted from Fig.~4 in Ref.~\cite{plenio01b}.)}\label{szilard}
\end{center}
\end{figure}

The process employed by Szilard's engine is schematically depicted in Fig.~\ref{szilard}. A
chamber of volume $V$ contains a gas, which consists of a single molecule (Fig.~\ref{szilard}(a)).
As a first step of the process, a thin, massless, adiabatic partition is inserted into the chamber
quickly to divide it into two parts of equal volumes. The demon measures the position of the
molecule, either in the right or in the left side of the partition (Fig.~\ref{szilard}(b)). The
demon records this result of the measurement for the next step. Then, he connects a load of a
certain mass to the partition on the side where the molecule is supposed to be in, according to
his recorded result of the previous measurement (Fig.~\ref{szilard}(c)). Keeping the chamber at a
constant temperature $T$ by a heat bath, the demon can let the gas do some work $W$ by quasistatic
isothermal expansion (the partition now works as a piston). The gas returns to its initial state,
where it now occupies the whole volume $V$, when the partition reaches the end of the chamber.
During the expansion, heat $Q$ is extracted from the heat bath and thus $W=Q$ as it is an
isothermal process. Hence, Szilard's engine completes a cycle after extracting heat $Q$ and
converting it to an equal amount of mechanical work.

As the gas is expanded isothermally \footnote{The load should be varied continuously to match the
pressure so that the expansion be a quasistatic and reversible process, and this enables the
pressure to be expressed as $p=kT/V$.}, the amount of extracted work $W$ is
$kT\int_{V/2}^{V}V^{-1}dV=kT\ln2$. An immediate question here might be if it is appropriate to
assume the one-molecule gas as a normal ideal gas in discussing thermodynamic/statistical
properties. To fill this conceptual gap, we can consider an ensemble of one-molecule gases. Then
by taking averages over the ensemble we can calculate various quantities as if it is an ideal gas
with a large number of molecules. In a sense, this can been seen as the origin of the intersection
between thermodynamics and information theory: looking at the (binary) position of the molecule
leads to its `dual' interpretations, i.e., in terms of thermodynamics and information theory.

Naturally, the factor $kT\ln2$ appears often in the following discussions on thermodynamic work,
so we will take it as a unit and call it `1 bit' when there is no risk of confusion\footnote{By
including the temperature $T$, we hereafter use the same unit `bit' for both entropy and
thermodynamical work.}. This will be especially useful when we coordinate discussions of the
information theoretic `bit' with the thermodynamic work.

The demon apparently violates the second law. As a result of the perfect conversion of heat $Q$
into work $W$, the entropy of the heat bath has been reduced by $Q/T=W/T=k\ln2$. According to the
second law, there must be an entropy increase of at least the same amount somewhere to compensate
this apparent decrease. Szilard attributed the source of the entropy increase to measurement. He
wrote ``The amount of entropy generated by the measurement may, of course, always be greater than
this fundamental amount, but not smaller" \cite{szilard29}. He referred to $k\ln2$ of entropy as a
fundamental amount well before Shannon founded information theory in 1948
\cite{shannon48,shannon49}. Although he regarded the demon's memory as an important element in
analyzing his one-molecule engine, Szilard did not reveal the specific role of the memory in terms
of the second law. Nevertheless, his work is very important as it was the first to identify the
explicit connection between information and physics.

\subsection{Temporary solutions to the paradox}\label{temp_solution}
As Szilard did, many generations believed for decades that the paradox of Maxwell's demon could be
solved by attributing the entropy increase to measurement. Noteworthy examples include those by
Brillouin Ref.~\cite{brillouin51} and Gabor \cite{gabor61}. They considered light to measure the
speed of the molecules and (mistakenly) assumed this to be the most general measurement setting.
Inspired by the work of Demers \cite{demers44}, who recognized in the 1940s that a high
temperature lamp is necessary to illuminate the molecules so that the scattered light can be
easily distinguished from blackbody radiation, Brillouin showed that information acquisition via
light signals is necessarily accompanied by an entropy increase, which is sufficient to save the
second law \cite{brillouin51}. Interestingly, in his speculation, Brillouin linked the
thermodynamic and the information entropies directly. Information entropy is a key function in the
mathematical theory of information, which was founded by Shannon only a few years before
Brillouin's work, and although its logical origin is very different from thermodynamics, Brillouin
dealt with two entropies on the same footing by putting them in the same equation to link the gain
of information with the decrease of physical entropy. This led to the idea of \textit{negentropy},
which is a quantity that behaves oppositely to the entropy\footnote{The idea of \textit{negative
entropy} itself was introduced by Schr\"{o}dinger to discuss living systems that keep throwing
entropy away to the environment. It was renamed as negentropy by Brillouin, who associated it with
information.} \cite{brillouin59,schroedinger44}. The negentropy is usually defined as the
difference between the maximum possible entropy of a system under a given condition and the
entropy it actually has, i.e. $N:=S_\mr{max}-S$.

Brillouin distinguished two kinds of information, \textit{free} and \textit{bound}. Free
information $I_f$ is an abstract and mathematical quantity, but not physical. Bound information
$I_b$ is the amount of information that can be acquired by measurement on a given physical system.
Thus, roughly speaking, free information is equivalent to (abstract) knowledge in our mind and
bound information corresponds to the information we can get about a physical system, which encodes
the information to be sent or stored. Bound information is then subject to environmental
perturbations during the transmission. When the information carrier is processed at the end of the
channel, it is transformed into free information. In Brillouin's hypothesis, the gain in bound
information by measurement is linked to changes in entropy in the physical system as
\begin{eqnarray}\label{bound_info}
\Delta I_b &=& I_b^{\mbox{\scriptsize{post-meas}}}-I_b^{\mbox{\scriptsize{pre-meas}}} \nonumber
\\
&=& k(\ln P_{\mbox{\scriptsize{pre-meas}}}-\ln
P_{\mbox{\scriptsize{post-meas}}}) \nonumber \\
&=& S_{\mbox{\scriptsize{pre-meas}}}-S_{\mbox{\scriptsize{post-meas}}}>0,
\end{eqnarray}
where $P_{\mbox{\scriptsize{pre-meas}}}$ and $P_{\mbox{\scriptsize{post-meas}}}$ denote the
numbers of possible states of the physical system before and after the measurement, and similarly
$S_{\mbox{\scriptsize{pre-meas}}}$ and $S_{\mbox{\scriptsize{post-meas}}}$ are the entropies of
the system \footnote{Equal probabilities for $P_{\mbox{\scriptsize{pre-meas}}}$ (or
$P_{\mbox{\scriptsize{post-meas}}}$) possible states are assumed.}. The conversion coefficient
between physical entropy and bound information is chosen to be Boltzmann's constant to make the
two quantities comparable in the same units. Equation (\ref{bound_info}) means that gaining bound
information decreases the physical entropy. This corresponds to the process (a) to (b) in
Fig.~\ref{szilard}.

As bound information is treated with the physical entropy of the system on the same basis, the
second law needs to be expressed with bound information as well as the physical entropy. If no
information on the physical system is available initially, that is, if $I_b^\mr{initial}=0$, the
final entropy of the system after obtaining (bound) information $I_b$ is $S_f=S_i-I_b$. The second
law of thermodynamics says that in an isolated system the physical entropy does not decrease
\footnote{Although, at first sight, taking $\Delta S:=S_f-S_i$ seems more natural to express the
second law, the subscripts only represent the state either before or after a measurement that
provides us with information on the system (bound information), not a physical time evolution.
Therefore the change in entropy due to physical evolution should be written as $\Delta S_f$.}:
$\Delta S_f\ge 0$. By using the change in negentropy $\Delta N:=-\Delta S$, the second law may now
be written as
\[
\Delta S_f=\Delta(S_i-I_b)=\Delta S_i-\Delta I_b=-\Delta N_i-\Delta I_b\ge 0,
\]
which means
\begin{equation}\label{ineq_negentropy}
\Delta(N_i+I_b)\le 0.
\end{equation}
Naturally, if there is no change in the information available to us, that is, $\Delta I_b=0$,
Eq.~(\ref{ineq_negentropy}) is nothing but the standard inequality for entropy, $\Delta S\ge 0$.
However, in Eq.~(\ref{ineq_negentropy}), information is treated as part of the total entropy, and
it states that the quantity (negentropy + information) never increases. This is a new
interpretation of the second law of thermodynamics, implied by Brillouin's hypothesis.

Following Brillouin's hypothesis, Lindblad compared the entropy decrease in the system with the
information gain an observer can acquire \cite{lindblad74}. He analyzed measurements of
thermodynamic quantities in a fluctuating system and showed that the information gain by the
observer is greater than or equal to the entropy reduction in the system. Hence, the total entropy
never decreases, as expected.

Brillouin's idea of dealing with information and physical entropy on an equal basis has been
widely accepted. All discussions below about the physical treatment of information processing
tacitly assume this interpretation, which presupposes the \textit{duality} of entropy, i.e. both
information theoretic and thermodynamic aspects.

\section{Exorcism of Maxwell's demon: Erasure of classical information encoded in classical states}\label{exorcism}

Although the exorcism of Maxwell's demon by attributing an entropy increase to the acquisition of
information had been widely accepted by physicists for more than a decade, the demon turned out to
have survived until Landauer and Bennett put an end to the demon's life by reconsidering the role
of ``memory", which Szilard barely overlooked. Landauer examined the process of erasure of
information, introducing a new concept of ``logical irreversibility" \cite{landauer61}.

Indeed, O. Penrose independently discovered essentially the same result about information erasure
as Landauer's. Penrose argued, in his book published in 1970, \textit{Foundations of Statistical
Mechanics}, that the paradox of Maxwell's demon could be solved by considering the entropy
increase due to memory erasure. This was even earlier than Bennett's 1982 analysis of the demon,
however, it was left virtually unnoticed by physicists. Penrose's treatment was rather abstract
and it did not go as far as Bennett's work, which investigated the possibility of measurement with
arbitrarily little entropy increase. Here we focus on the viewpoint by Landauer and Bennett.

Since information processing must be carried out by a certain physical system, there should be a
one-to-one correspondence between logical and physical states. Logical states may be described as
an abstract set of variables on which some information processing can be performed. Then, a
reversible logical process, which means an injective (one-to-one) mapping for logical states,
corresponds to a reversible physical process. By implicitly assuming a correspondence between
logical and physical entropies, as Brillouin proposed, this implies that a reversible logical
process can be realized physically by an isentropic process, i.e. an entropy-preserving process.

However, a logically irreversible process is non-injective, i.e. many-to-one, mapping. Such a
process does not have a unique inverse as there may be many possible original states for a single
resulting state. The key here is that memory erasure is a logically irreversible process because
many possible states of memory should be set to a single fixed state after an erasing procedure.
It is impossible to determine the state prior to erasure without the aid of further information,
such as the particular task of a computer programme or knowledge about the states of other memory
registers that are correlated to the memory in question. This certain fixed state after erasure is
analogous to a ``white" or ``blank" sheet of paper, on which no information is recorded. After
erasing stored information, the state of memory should be in one specific state, in order not to
carry any information (by definition of erasure). We will refer to the specific state after
erasure as a \textit{standard state}.

In terms of physical states, a logically irreversible process reduces the degrees of freedom of
the system, which implies a decrease in entropy. In order for this process to be physically
legitimate, the energy must be dissipated into the environment. Landauer then perceived that
logical irreversibility must involve dissipation, hence erasing information in memory entails
entropy increase (in the environment). This point will be the final sword to exorcize Maxwell's
demon and is referred to as Landauer's \textit{erasure principle}.

Another important observation regarding the physics of information was given by Bennett
\cite{bennett82}. He illustrated that measurement can be carried out reversibly, i.e. without any
change in entropy, provided the measuring apparatus is initially in a standard state, so that
recording information in the memory does not involve the erasure of information previously stored
in the same memory. The rough reason for this is that measurement can be regarded as a process
that correlates the memory with the system (in other words, a process that copies the memory state
to another system in a standard state), which can be achieved reversibly, at least in principle.

\begin{figure}
\begin{center}
\includegraphics[scale=0.35]{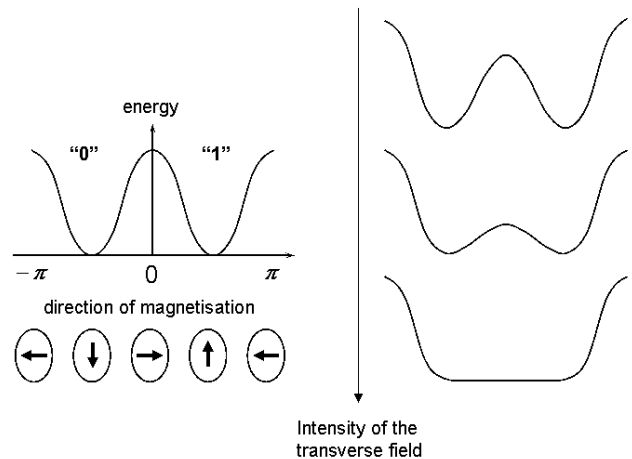}
\caption{A potential energy for a binary memory whose state is represented by the direction of the
magnetization. When there is no external transverse magnetic field the memory is stable in one of
the potential wells, which correspond to ``0" and ``1" of recorded information. The transverse
field lowers the height of the barrier at the centre. At a certain point, the profile of the
potential becomes bath-tub-like shape with a flat bottom, where the direction of the magnetization
is very sensitive to the longitudinal field component. (Adapted from Ref.~\cite{bennett82}.)}
\label{magnetic}
\end{center}
\end{figure}

Bennett exemplified the reversible correlating process by a one bit memory consisting of an
ellipsoidal piece of ferromagnetic material. The ferromagnetic piece is small enough so that it
consists of only a single domain of magnetization. The direction of the magnetization represents
the state of the memory. Suppose that there is a double well potential with respect to the
direction of the magnetization in the absence of an external field: Parallel and antiparallel to
the major axis of the ellipsoid are the most stable directions. The central peak between the two
wells is considerably higher than $kT$ in the absence of an external magnetic field, so that
thermal fluctuations do not allow the state to climb over the peak. Figure \ref{magnetic} shows a
sketch of the potential that was illustrated in Ref.~\cite{bennett82}.

Two minima of the potential represent the state of memory, either ``0" or ``1", and the blank
memory is assumed to be in one standard state, e.g. ``0", before information is copied onto it
from another memory. Let us consider the process that correlates the state of a blank memory $B$
with that of a memory $A$, which is the subject of measurement. This can be achieved by
manipulating the shape of the potential for the blank memory as follows. By applying a transverse
external magnetic field, the peak of the central barrier becomes lower. At a certain intensity of
the field, there will be only a single bath-tub-like flat bottom, i.e. the state of $B$ becomes
very sensitive to a weak longitudinal component of the field. The memory $A$ is located so that
its magnetization can cause a faint longitudinal field at the position of the memory $B$. Then
because of $B$'s sensitivity to such a field, the state of $A$ can be copied to the memory $B$
with arbitrarily small (but nonzero) energy consumption. Removing the transverse field completes
the correlating process. The crux of the physics here is that this process can be reversed by
using the perturbation from another reference memory, which is in the standard state.

Now let us focus on the erasure of information. Since measurement can be done virtually without
energy consumption, it is the dissipation due to the erasure process that compensates the entropy
decrease induced by Maxwell's demon in Szilard's model. The physical system for the demon's memory
can be modelled as a one-molecule gas in a chamber of volume $V$, which is divided into two parts,
the left ``$L$" and the right ``$R$", by a partition. The demon memorizes the measurement result
by setting the position of the molecule in this box. If the molecule in Szilard's engine may be
found in the left and the right sides with equal probability, i.e. 1/2, then the minimum amount of
work that needs to be invested and dissipated into the environment is $kT\ln2$.

\begin{figure}
\begin{center}
\includegraphics[scale=0.66]{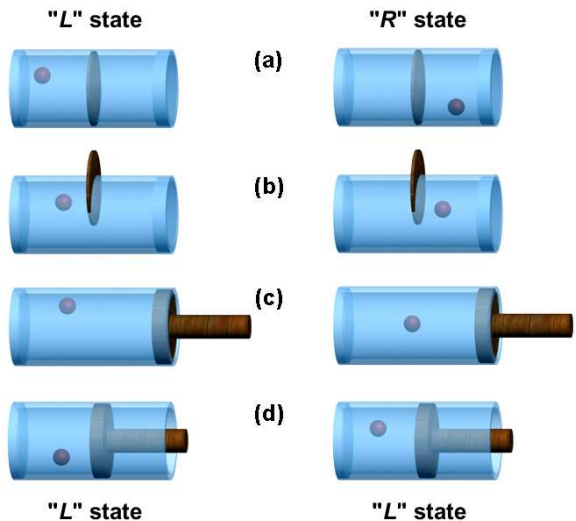} 
\caption{Thermodynamic process to erase information. A binary information is stored in a vessel as
the position of the molecule, either $L$ or $R$. A common procedure for both initial states, i.e.
removing the partition and halving the whole volume by an isothermal compression towards the
standard state $L$, completes the erasure. (Adapted from Fig.~3 in Ref.~\cite{plenio01b}.)}
\label{erasure_gas}
\end{center}
\end{figure}

The actual process is as follows. The molecule is in either $L$ or $R$, depending on the
information it stores (Fig.~\ref{erasure_gas}(a)). To erase the stored information, first, we
remove the partition dividing the vessel at the centre (Fig.~\ref{erasure_gas}(b)). Second, insert
a piston at the right end (Fig.~\ref{erasure_gas}(c)), when the standard memory state is $L$, and
push it towards the left isothermally at temperature $T$ until the compressed volume becomes $V/2$
(Fig.~\ref{erasure_gas}(d)). The resulting state is $L$ for both initial states and the
information is erased. It is worth noting that the erasing process should not depend on the
initial state of the memory. The ``$R$" state in Fig.~\ref{erasure_gas}(a) may be transferred to
``$L$" state by simply moving the region of volume $V/2$ to the left. However, in this case, the
operator of the piston needs to observe the position of the molecule and this action requires
another memory. Thus, the erasure process should be independent of the initial memory state. The
work invested to compress the volume from $V$ to $V/2$ is $W_\mathrm{erasure}=kT\ln2$ and this is
dissipated as heat into the environment, increasing its entropy by $k\ln2$, as Landauer argued. As
there is no wasted work (in the sense that all invested work is converted into heat to increase
the entropy of the environment), $kT\ln2$ is the minimum amount of work to be consumed for
erasure.

\begin{figure}
\begin{center}
\includegraphics[scale=0.38]{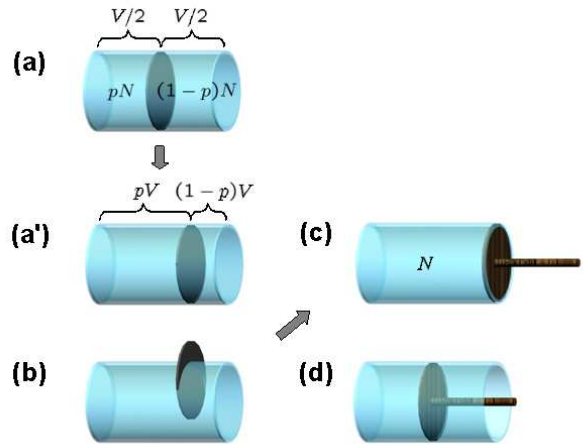} 
\caption{Erasure process for an unbalanced probability distribution. The only difference from the
case of balanced distribution (Fig.~\ref{erasure_gas}) is the expansion from (a) to (a'), which
gives us $H(p)$ bits of work.} \label{erasure_gas2}
\end{center}
\end{figure}

If there is a biased tendency in the frequency of appearance of a particular memory state, say
$L$, how much would the erasure work be? The answer is simple: the erasure work is proportional to
the amount of information stored, thus $W_\mathrm{erasure}=kT\ln2H(p)$, where $p$ is the
probability for the molecule to be in the $L$ state and
\begin{equation}
H(p)=-p\log p-(1-p)\log(1-p)
\end{equation}
is the (binary) Shannon entropy. Throughout this article, $\log$ denotes logarithms of base 2. The
reason can be explained by a process depicted in Fig.~\ref{erasure_gas2}. The unbalanced tendency
between $L$ and $R$ is expressed by the numbers of molecules in the $L$ and the $R$ regions. As we
consider only an ideal gas (with no interactions between molecules), this scenario does not change
the discussion at all if we average the erasure work at the end. Since removing the partition at
the beginning allows the gas an undesired irreversible adiabatic expansion/compression, we first
let the gases in both parts expand/contract isothermally by making the partition movable without
friction (Fig.~\ref{erasure_gas2}(a) to (a')). During this process, the gases exert work towards
the outside. Letting $p_L$, $p_R$ and $V_L$ denote the pressure in the left region, that in the
right region, and the volume of the region on the left of the partition, respectively, we can
write the work done by gases as
\begin{eqnarray}\label{exp-comp}
W^\prime &=& \int_{V/2}^{pV}(p_L-p_R)dV_L \nonumber \\
&=& NkT\int_{V/2}^{pV}\left(\frac{p}{V_L}-\frac{1-p}{V-V_L}\right)dV_L \nonumber \\
&=& NkT (\ln2+p\ln p+(1-p)\ln(1-p)) \nonumber \\
&=& NkT\ln2(1-H(p)).
\end{eqnarray}
As the pressures in the left and the right are equal, this is the same situation as in
Fig.~\ref{erasure_gas}(a). Hence, at least $NkT\ln2$ of work needs to be consumed to set the
memory to the standard state (Fig.~\ref{erasure_gas2}(c) to (d)). As a whole, we invested
$W_\mathrm{erasure}=kT\ln2-W^\prime=kT\ln2 H(p)$ of work per molecule.

Maxwell's demon is now exorcized. The entropy decrease, or the equivalent work the demon could
give us, should be completely consumed to make his memory state come back to its initial state.
The state of the whole system, consisting of the heat engine and the demon, is restored after
completing a thermodynamic cycle, without violating the second law.

\section{Other `derivations' of the erasure entropy}
We have focused on the one-molecule gas model so far, however, Landauer's erasure principle holds
regardless of specific physical models. In order to see its generality with some concrete
examples, we now briefly review two particularly interesting papers, one by Shizume
\cite{shizume95} and the other by Piechocinska \cite{piechocinska00}.

Shizume used a model of memory whose state was represented by a particle having Brownian motion in
a time-dependent double well potential. Assuming the random force $F_R(t)$ to be white and
Gaussian satisfying $\langle F_R(t_1) F_R(t_2) \rangle = 2m\gamma T \delta (t_1-t_2)$, the motion
of the distribution function $f(x,u,t)$ of the particle in the position ($x$) and velocity ($u$)
space can be described by the Fokker-Planck equation. Shizume then compared $\dot{Q}$ and
$TdS/dt$, i.e. the ensemble average of the energy given to the particle from the environment per
unit time, and the change in the entropy of the whole system per unit time multiplied by the
temperature. The entropy $S$ is the Shannon entropy of continuous distribution, defined by
\begin{equation}
S:=-k\int_\infty^\infty dx\,du f(x,u,t)\,\ln f(x,u,t).
\end{equation}
With the help of the Fokker-Planck equation concerning $f(x,u,t)$, one arrives at the relation,
\begin{equation}
\dot{Q}\le T\frac{dS}{dt},
\end{equation}
from which we obtain the lower bound of the energy dissipated into the environment between times
$t_i$ and $t_f$ as
\begin{equation}\label{qdot_bound}
\Delta Q_\mathrm{out}(t_i, t_f) = \int_{t_i}^{t_f} (-\dot{Q})\, dt \,\ge\, T[S(t_i)-S(t_f)].
\end{equation}

Equation (\ref{qdot_bound}) gives us the lower bound of the heat generation due to the process
that erases $H(p)$ bits of information. As we expect, the lower bound is equal to $kT\ln2 H(p)$.
Clearly, this derivation does not use the second law.

It is clear in Shizume's derivation that the entropy increase due to the erasure is independent of
the second law. Hence it is immune to a common criticism against the erasure principle that it is
trivially the same as the second law because the second law is used in its derivation. However,
the above description assumes only a specific physical model and thus a more general model might
be desirable. This was done by Piechocinska, who analyzed the information erasure in a quantum
setting as well as in classical settings \cite{piechocinska00}.

The key idea in her results is to make use of a quantity $\Gamma$, which was introduced by
Jarzynski in the context of nonequilibrium thermodynamic processes \cite{jarzynski99}. In a
classical setting, $\Gamma$ is defined by
\begin{eqnarray}\label{gamma_jarzynski}
\Gamma(\zeta^0,\zeta^\tau) &=& -\ln [\rho_f (x^\tau,p^\tau)]+\ln [\rho_i (x^0,p^0)] \nonumber \\
& & +\beta \Delta E(\vec{x}_T^{\,0},\vec{p}_T^{\;0},\vec{x}_T^{\,\tau},\vec{p}_T^{\;\tau}),
\end{eqnarray}
where $\zeta=(x,p,\vec{x}_T,\vec{p}_T)$ is a set of positions and momenta of the degrees of
freedom that describe the (memory) system and the heat bath ($T$), respectively. The superscripts,
0 and $\tau$, are the initial and final times of the erasure process. $\rho_i$ and $\rho_f$ are
the distribution functions of the particle representing the `bit' in a double well potential, and
are assumed to be in the canonical distribution: erasing information is expressed by the form of
$\rho_f$ so that it takes nonzero values only in one of the two regions, i.e. either of those for
`0' and `1', which corresponds to the `L' state in Fig.~\ref{erasure_gas}. $\Delta E$ is the
change in the internal energy of the heat bath and $\beta=(kT)^{-1}$.

The entropy increase due to erasure can be obtained by first calculating the statistical average
over all possible trajectories $\zeta$. We then have $\langle e^{-\Gamma}\rangle=1$, which in turn
implies $-\langle \Gamma \rangle \le 0$ by the convexity of the exponential function. Substituting
the expressions for $\rho_i$ and $\rho_f$ (the canonical distributions) to $\Gamma$ leads to an
inequality
\begin{equation}\label{pie01}
\ln 2 \le \beta \langle \Delta E \rangle.
\end{equation}
As $\Delta E$ is the change in the internal energy of the heat bath, it includes the heat
dissipated into the bath as well. Thus the conservation of energy can be written as $W=\Delta E +
\Delta E_\mathrm{system}$, where $W$ is the work done on the system and the heat bath, and $\Delta
E_\mathrm{system}$ is the change in the internal energy of the system. Due to the symmetry of
$\rho_i$ and $\rho_f$, $\Delta E_\mathrm{system}$ vanishes when averaged over, therefore we now
have
\begin{equation}\label{pie02}
kT\ln 2 \le \langle W \rangle,
\end{equation}
which is equivalent to Landauer's erasure principle.

Piechocinska applied a similar argument to the quantum case, i.e. the erasure of classical
information stored in quantum states. The state of the bit can be reset after some interaction
with the heat bath, which is initially in thermal equilibrium. By assuming that the bath decoheres
into one of its energy eigenstates due to the interaction with an external environment, which may
be much larger than the bath, we can deal with the heat dissipation (into the bath)
quantitatively. Then, the minimum work consumption can be found to be $kT\ln2$ after computing a
quantity corresponding to $\Gamma$ in Eq.~(\ref{gamma_jarzynski}). A related, but more general,
argument in a similar spirit has been also presented in \cite{kawai07}.

\section{Some interesting implications of the second law}
\subsection{Wave nature of light from the second law}\label{wave_and_2ndlaw} 
\begin{figure}
\begin{center}
\includegraphics[scale=1]{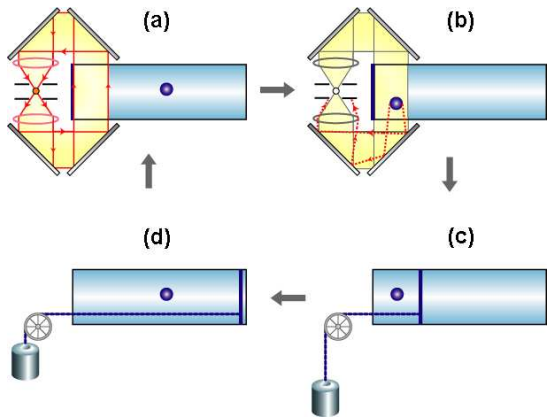}
\caption{The one-molecule heat engine considered by Gabor to show that the light needs to behave
like wave. When the molecule comes into the illuminated region, a piston is automatically inserted
and the 'gas' expands isothermally to extract work from the heat bath. This process could in
principle be repeated infinitely, converting infinite amount of heat into mechanical work, if the
light had only a particle-like nature. (Adapted from Fig.~7 in Ref.~\cite{gabor61}.)}
\label{gabor_engine}
\end{center}
\end{figure}

Let us make a detour
to another interesting and well-elaborated implication of the second law, which was argued by
Gabor. He studied Brillouin's analysis of Maxwell's demon about detecting a molecule by light
signals further \cite{gabor61}. Gabor considered a one-molecule heat engine, a part of which is
illuminated by an incandescent light to detect the molecule wandering into this region
(Fig.~\ref{gabor_engine}). The detection of the molecule can be done by photosensitive elements
that are placed around the light path, so that any scattered weak light will hit one (or more) of
them. As soon as the molecule is found by detecting the scattered light, a piston is inserted at
the edge of the illuminated region. Then by isothermal expansion the gas exerts mechanical work.
The same process is repeated when the molecule wanders into the illuminated region again. This is
a perpetuum mobile of the second kind as it continues to convert heat from a heat bath to
mechanical work. Gabor found that the second law is vulnerable if the light intensity can be
concentrated in a well defined region and made arbitrarily large compared with the background
blackbody radiation. He then deduced that this is impossible because light behaves both as waves
and as a flux of particles. In other words, according to Gabor's argument, the second law implies
the wave nature of light. This is an interesting implication of the second law in its own right,
as there seems to be no direct link between thermodynamics and the nature of light.

However, we now know that this interpretation is wrong. Even if the light behaves like particles,
Gabor's engine does not violate the second law. The solution to this apparent paradox is also the
erasure principle.

By detecting the molecule and extracting work subsequently, the whole system stores $H(p)$ bits of
information, where $p$ is the probability of finding the molecule in the illuminated region. Let
us assume for simplicity that the sampling frequency is low enough, compared with the time
duration necessary for the molecule to travel through the illuminated region \footnote{Otherwise,
the molecule can always be detected near the edge of the illuminated region. This region can then
be made as thin as the size of the molecule to maximize the work-extracting efficiency.}. Then,
$p$ can also be interpreted as the ratio between the volume of the illuminated region and the
whole volume of the chamber. This information is about the occurrence of the work extraction and
is stored in the mechanism that resets the position of the piston after the extraction. The piston
forgets the previous action, but the resetting mechanism does not. Thus the whole process is not
totally cyclic, though it should be so to work as a perpetuum mobile. Because Gabor's engine is
activated with probability $p$, the engine stores $H(p)$ bits on average. While $kT\ln2 H(p)$ bits
of work are needed to erase this information to make the whole process cyclic, we gain only
$-kTp\ln p$ of work, which is smaller than the erasure work, from this process, hence there is no
violation of the second law.

\subsection{Gibbs paradox and quantum superposition principle}\label{superposition_and_gibbs}
Suppose a gas chamber of volume $V$ that is divided into two half regions by a removable
partition. Each half region is filled with a dilute ideal gas at the same pressure $P$. We now
consider the entropy increase that occurs when we let gases expand into the whole volume $V$ by
removing the partition. If the gas in one region (e.g., the left side) is different from the gas
in the other (the right side), then the entropy increase due to the mixing is $kN\ln2$, where $N$
is the total number of molecules in the chamber. On the other hand, if the gases in two regions
are identical, no thermodynamic change occurs and thus the entropy is kept constant. This
discontinuous gap of the entropy increase with respect to the similarity of two gases is called
the `Gibbs paradox'. Land\'{e} dealt with this problem and `derived' the wave nature of physical
state as well as the superposition principle of quantum mechanics \cite{lande52}. This is not only
an interesting work in the sense that it attempted to link thermodynamics and quantum mechanics,
but useful to introduce the idea of semipermeable membranes that we will use as a tool in later
sections. Although there are a number of papers on the Gibbs paradox other than Land\'{e}'s, we
feel that going into these is out of the scope of this brief review. Interested readers may refer
to, for example, Refs.~\cite{jaynes92,lyuboshitz70,levitin93,allahverdyan06}.

For convenience of the following discussion, we use the extractable work, i.e. the Helmholtz free
energy, as it is equal to the entropy change (times temperature) in isothermal processes. The
semipermeable membranes we introduce here are a sort of filters that distinguish the property of
gases, i.e. the nature of molecules, and let one (or more) particular property of gas go through
it. In other words, a semipermeable membrane is transparent to one type of gas, but totally opaque
to other types of gases: Each membrane thus can be characterized with the property of the gas it
lets go through. These are essentially the same as what von Neumann considered in his discussion
to define the entropy of a quantum state, or \textit{von Neumann entropy} \footnote{Von Neumann
defined the entropy $S$ of a quantum state $\rho$ by a simple thermodynamic consideration
\cite{vonneumann}. Suppose a vessel is filled with an ideal gas, every molecule of which is in the
state $\rho$. One now decomposes the gas into the set of gas components, each of which is in a
pure state $\ket{\psi_i}$, with the semipermeable membranes. The entropy $S(\rho)$ is defined (up
to a constant factor) as the minimal thermodynamic entropy increase in the environment that is
necessary to transform the initial state to the final state, where every molecule is in the same
pure state and is distributed uniformly over the whole vessel. The zero entropy for any pure state
is postulated.}, and they were scrutinized by Peres and were shown to be legitimate quantum
mechanically \cite{peres90,peresbook}. In Land\'{e}'s argument, however, quantum mechanics is not
assumed from the outset.

Land\'{e} postulated the continuity of the entropy change in reality. To bridge the gap between
the `same' and `different' gases, he introduced a fractional likeness, which is quantified as
$q(A_i,B_j)$, between two states $A_i$ and $B_j$. Here $A$ (or $B$) represents a certain
`property' or `observable' and the indices are values of $A$ ($B$) with which we can distinguish
them completely. For simplicity, we assume all observables can take only discrete values when
measured. Two states are completely different when $q=0$ and they are identical when $q=1$, and
different values of the same property are perfectly distinguishable by some physical means, thus
$q(A_i,A_j)=\delta_{ij}$. Now suppose that a semipermeable membrane that is opaque to $A_i$ but
transparent to $A_j (j\neq i)$, to which we will refer as the membrane $M_{A_i}$, is placed in a
gas whose property is $B_k$. Then the membrane will reflect a fraction $q=q(A_i,B_k)$ of the gas
and pass the remaining fraction $1-q$ as a result of the fractional likeness between $A_i$ and
$B_k$. Another consequence of the membrane is that the molecules that are reflected by $M_{A_i}$
need to change their property from $B_k$ to $A_i$ and similarly the other molecules become $A_j$
with probability $q(A_j,B_k)$ \footnote{This means that the membrane $M_{A_i}$ performs a
measurement about the property $A$ just before the molecule hits it and the molecule's
post-measurement property will become $A_j$ with probability $q(A_j,B_k)$.}, in order not to
change the state of molecules by a subsequent application of another $M_{A_i}$. In what follows,
we will identify the term `property' with `state', although it still does not necessarily mean a
quantum state.

This solution to the Gibbs paradox -- the introduction of fractional likeness between states --
leads, according to Land\'{e}, to the wave-function-like description of state. A rough sketch of
his idea is as follows. First, we write down the transition probabilities between different states
in a matrix form
\begin{equation}\label{q_matrix}
\left[\begin{array}{ccc} q(A_1,B_1) & q(A_1,B_2) & \cdots \\ q(A_2,B_1) & q(A_2,B_2) & \cdots \\
\cdots & \cdots & \cdots \end{array}\right].
\end{equation}
Naturally, the sum of each row or column is always unity because a state must take one of the
possible values in any measured property. Similar matrices should be obtained for $q(B,C)$,
$q(A,C)$, etc., and we expect a mathematical relation between these matrices, for instance, such
as $q(A_i,C_k)\stackrel{\mbox{\small{?}}}{=}\sum_j q(A_i,B_j)q(B_j,C_k)$. A consistent
mathematical expression can be obtained by considering a matrix $\psi(A,B)$ whose $(i,j)$-th
elements are given as $\sqrt{q(A_i,B_j)}e^{i\varphi}$ with arbitrary phase $\varphi$, i.e. a
matrix whose rows and columns can be regarded as a vector of unit norm, thus $\psi$ for different
pairs of properties are connected with an orthogonal (unitary) transformation. The arbitrariness
for the phase is restricted by the condition $\psi(A_i,A_j)=\sum_k
\psi(A_i,B_k)\psi(B_k,A_j)=\delta_{ij}$. Identifying $\psi(A_i,B_j)=\sqrt{q}e^{i\varphi}$ with a
complex probability amplitude for the transition from $B_j$ to $A_i$ induced by the membranes, we
see a superposition rule $\psi(A_i,C_k)=\sum_j\psi(A_i,B_j)\psi(B_j,C_k)$. Then Land\'{e} claims
that \textit{``the introduction of complex probability amplitudes $\psi$ subject to the
superposition rule is inseparably linked to the admission of fractional likenesses $q$."}

\begin{figure}
\begin{center}
\includegraphics[scale=0.8]{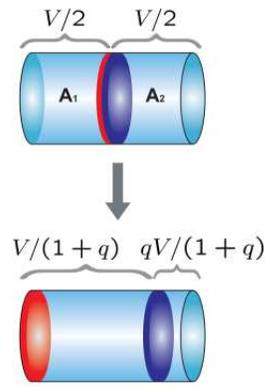}
\caption{A possible configuration to confirm the continuity of the extractable work. The left
(right) hand side of a chamber is filled with an $A_1$ $(B_1)$-gas. Two membranes that distinguish
$A_1$ and $A_2$ are used to extract work. The membrane on the left lets the $A_1$ gas pass it
through freely, but reflects the $A_2$ gas completely. The other membrane works in the opposite
manner. Since the $B_1$ gas changes its state into $A_1$ with probability $p$ when measured by the
membrane, the right membrane does not reach the right end of the chamber by a (quasi-static)
isothermal expansion.} \label{lande_superposition}
\end{center}
\end{figure}

To confirm the continuity of the extractable work (or the entropy increase) due to the mixing of
two gases, let us look at a chamber, a half of which is filled with dilute gas $A_1$ and the other
half with $B_1$ as in Fig.~\ref{lande_superposition}. The number of gas molecules is $N/2$ each.
Let us also assume that both $A$ and $B$ are a two-valued property. If we use two membranes that
distinguish the state $A_1$ and $A_2$, the work by gases will be smaller than $NkT\ln2$ because a
fraction of $B_1$ becomes $A_1$ with a certain probability $q$. The work done by the gases is
given as $W=(NkT/2)[2\ln2+q\ln q-(1+q)\ln (1+q)]$ and $W$ decreases smoothly from $NkT\ln2$ (when
$q=0$, for perfectly distinct gases) to 0 (when $q=1$, identical gases), therefore no
discontinuous entropy change. Note that this choice of membranes is not optimal to maximize the
amount of extractable work and we will look at this process in more detail in Section
\ref{holevo_from_2ndlaw}.

\subsection{Quantum state discrimination and the second law}\label{distinguishability_and_2ndlaw}
As we have seen in the previous section, in his attempt to solve the Gibbs paradox, Land\'{e}
deduced that a thermodynamic speculation in the form of the continuity principle could lead to the
partial likeness (or distinguishability) of states as a result of the wave nature of particles. On
the other hand, starting from the distinguishability issue of quantum states, Peres showed that if
it was possible to distinguish non-orthogonal quantum states perfectly then the second law of
thermodynamics would necessarily be violated \cite{peres90,peresbook}.

\begin{figure}
\begin{center}
\includegraphics[scale=0.8]{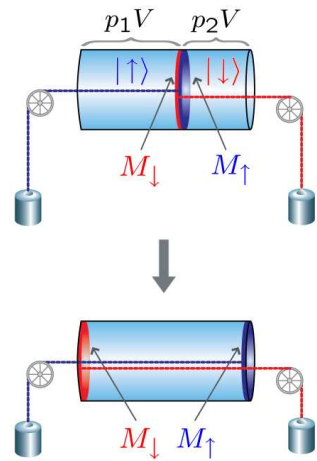}
\caption{The work-extracting process with semipermeable membranes. In the initial state (a), the
vessel is divided into two parts by an impenetrable opaque partition. The left side of the vessel,
whose volume is $p_1 V$, is occupied by the $\ket{\uparrow}$-gas, and the right side is filled
with $\ket{\!\downarrow}$-gas. By replacing the partition with two semipermeable membranes,
$M_{\uparrow}$ and $M_{\downarrow}$, we can extract $H(p_1)=-p_1\log p_1-p_2\log p_2$ bits of work
by isothermal expansion. The membranes reach the end of the vessel in the final state (b).}
\label{ex_membrane}
\end{center}
\end{figure}

As a background, let us consider an elementary work-extraction process using a collection of pure
orthogonal states. As shown in Fig.~\ref{ex_membrane}, a chamber of volume $V$ is partitioned by a
wall into two parts, one of which has a volume $p_1 V$ and the other has $p_2 V$, where
$p_1+p_2=1$. The vessel is filled with a gas of molecules whose (quantum) internal degree of
freedom is represented by, for example, a spin. Here it suffices to consider a gas of spin-1/2
molecules, e.g. a gas with spin up, i.e. $\ket{\!\uparrow}$, in the left region and a spin down
gas, $\ket{\!\downarrow}$, in the right.

Now we reintroduce semipermeable membranes, $M_{\uparrow}$ and $M_{\downarrow}$, that distinguish
the two orthogonal spins, $\ket{\!\uparrow}$ and $\ket{\!\downarrow}$. These are essentially the
same as what we have seen in Section \ref{superposition_and_gibbs} to consider the `fractional
likeness' of states. The membrane $M_{\uparrow}$ is completely transparent to the
$\ket{\!\downarrow}$ gas and completely opaque to the $\ket{\!\uparrow}$ gas. The other membrane
$M_{\downarrow}$ has the opposite property.

Suppose that the partition separating two gases is replaced by the membranes, so that
$M_{\uparrow}$ and $M_{\downarrow}$ face $\ket{\!\uparrow}$ and $\ket{\!\downarrow}$,
respectively, as in Fig.~\ref{ex_membrane}. Then, as in Section \ref{superposition_and_gibbs},
gases give us some work, expanding isothermally by contact with a heat bath of temperature $T$.
The total work extractable can then be computed as $W=-p_1\log p_1-p_2\log p_2=H(p_i)$, where
$H(p_i)=-\sum_i p_i\log p_i$ is the Shannon entropy of a probability distribution $\{p_i\}$.

\begin{figure}
\begin{center}
\includegraphics[scale=0.85]{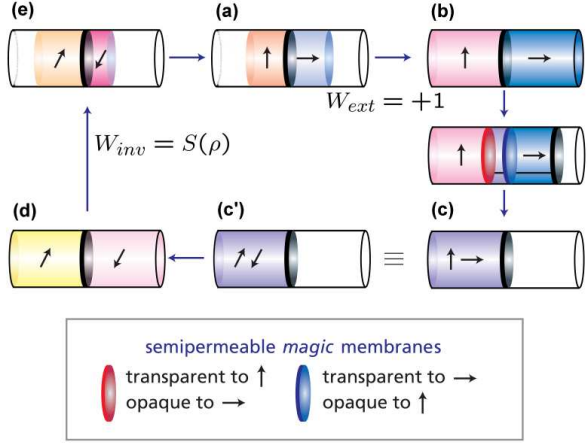}
\caption{A thermodynamic cycle given by Peres to show that distinguishing non-orthogonal quantum
states leads to a violation of the second law. The arrows indicate the directions of spin in the
Bloch sphere. The use of hypothetical semipermeable membranes, which distinguish non-orthogonal
states $\ket{\!\uparrow}$ and $\ket{\rightarrow}$ perfectly, in the step from (b) to (c) is the
key to violate the second law. (Modified from Fig.~9.2 in Ref.~\cite{peresbook}.)}
\label{peres_cycle}
\end{center}
\end{figure}

Let us look at Peres's process. The physical system we consider now is almost the same as the one
in Fig.~\ref{ex_membrane}, however, we now have two non-orthogonal states. Although gases consist
of photons of different polarizations in the original example by Peres, we consider spin-$1/2$
molecules to avoid the argument of the particle nature of light. The volume of the chamber here is
$2V$, and in the initial state, the gas of volume $V$ is divided into two equal volumes $V/2$ and
separated by an impenetrable wall (See Fig.~\ref{peres_cycle}(a)). The gas molecules in the left
side have a spin up $\ket{\!\uparrow}$, and those in the right side have a spin
$\ket{\!\rightarrow}=(\ket{\!\uparrow}+\ket{\!\downarrow})/\sqrt{2}$. Both parts have the same
number of molecules, $N/2$, thus the same pressure. The first step is to let gases expand
isothermally at temperature $T$ so that the entire chamber will now be occupied by them
(Fig.~\ref{peres_cycle}(b)). During this expansion, gases exert 1 bit of work ($=NkT\ln2$) towards
the outside, absorbing the same amount of heat from the heat bath.

In the second step, we introduce fictitious ``magic" membranes that can distinguish non-orthogonal
states. We replace the partition at the centre with these membranes and insert an impenetrable
piston at the right end of the vessel. The membrane $M_{\uparrow}^\prime$, which is transparent to
the $\ket{\!\rightarrow}$-gas but opaque to $\ket{\!\uparrow}$-gas, is fixed at the centre, while
the other one $M_{\rightarrow}^\prime$, which has the opposite property to $M_{\uparrow}^\prime$,
can move in the area on the left. Then, the piston inserted at the right end and
$M_{\rightarrow}^\prime$ is pushed towards the left at the same speed so that the volume and the
pressure of the $\ket{\!\rightarrow}$-gas in between the piston and the membrane
$M_{\rightarrow}^\prime$ will be kept constant (from Fig.~\ref{peres_cycle}(b) to (c)). Because of
the property of the membranes, this step can be achieved without friction/resistance, thus needs
no work consumption or heat transfer.

The gas in the vessel in Fig.~\ref{peres_cycle}(c) is a mixture of two spin states. The density
matrix for this mixture is
\begin{equation}\label{spin-mixture01}
\rho = \half \ket{\!\uparrow}\bra{\uparrow}+\half\ket{\rightarrow}\bra{\rightarrow} =
\frac{1}{4}\left(\begin{array}{cc} 3 & 1 \\ 1 & 1\end{array}\right)
\end{equation}
in the $\{\ket{\!\uparrow},\ket{\!\downarrow}\}$-basis. The eigenvalues of $\rho$ are
$(1+\sqrt{2}/2)/2=0.854$ and $(1-\sqrt{2}/2)/2=0.146$ with corresponding eigenvectors
$\ket{\!\nearrow}=\cos\frac{\pi}{8}\ket{0}+\sin\frac{\pi}{8}\ket{1}$ and
$\ket{\!\swarrow}=\cos\left(-\frac{3\pi}{8}\right)\ket{0}+\sin\left(-\frac{3\pi}{8}\right)\ket{1}$,
respectively.

Now let us replace the ``magic" membranes by ordinary ones, which discriminate two orthogonal
states, $\ket{\!\nearrow}$ and $\ket{\!\swarrow}$. The reverse process of (b)$\rightarrow$(c) with
these ordinary membranes separates $\ket{\!\nearrow}$ and $\ket{\!\swarrow}$ to reach the state
(d). Then, after replacing the semipermeable membranes by an impenetrable wall, we compress the
gases on the left and the right parts isothermally until the total volume and the pressure of the
gases become equal to the initial ones, i.e. those in state (a). This compression requires a work
investment of $-(0.854\log 0.854+0.146\log 0.146)=0.600$ bits, which is dissipated into the heat
bath. In order to return to the initial state (a) from (e), we rotate the direction of spins so
that the left half of the gas becomes $\ket{\!\uparrow}$ and the right half becomes
$\ket{\!\rightarrow}$. More specifically, we insert an opaque wall to the vessel to halve the
volume $V$ occupied by gases (the border between regions labelled A and B in
Fig.~\ref{peres_cycle}(e)). Rotations $\ket{\!\nearrow}\rightarrow\ket{\!\uparrow}$ in the region
A, $\ket{\!\nearrow}\rightarrow\ket{\!\rightarrow}$ in B and
$\ket{\!\swarrow}\rightarrow\ket{\!\rightarrow}$ in C, and a trivial spatial shift restore the
initial state (a). As rotations here are unitary transformations, thus an isentropic process, any
energy that has to be supplied can be reversibly recaptured. Alternatively, we can put the system
in an environment such that all these spin pure states are degenerate energy eigenstates. Hence,
we do not have to consider the work expenditure in principle when the process is isentropic.

Throughout the process depicted above and in Fig.~\ref{peres_cycle}, the net work gained is
$1-0.600=0.400$ bits. Therefore, Peres's process can complete a cycle that can withdraw heat from
a heat bath and convert it into mechanical work without leaving any other effect in the
environment. This implies that the second law sets a barrier to quantum state discrimination.

\subsection{Linearity in quantum dynamics}\label{linearity_of_QM}
Peres also showed that the second law should be violated if we admit nonlinear (time) evolution of
the quantum states \cite{peres89}. His proof is concise and is summarized below.

Let the state $\rho$ be a mixture of two pure states,
$\rho=p\ket{\phi}\bra{\phi}+(1-p)\ket{\psi}\bra{\psi}$, with $0<p<1$. By rewriting one of the
state vectors, say $\ket{\psi}$, as $\ket{\psi}=\sqrt{f}\ket{\phi}+\sqrt{1-f}\ket{\phi^\perp}$,
where $\ket{\phi^\perp}$ is a vector orthogonal to $\ket{\phi}$ and $f=|\braket{\phi}{\psi}|^2$,
$\rho$ can be written in a matrix form (in the 2-dimensional subspace that supports $\rho$) as
\begin{equation}\label{rho_matrix}
\rho=\left( \begin{array}{cc} p+f(1-p) & \sqrt{f(1-f)}(1-p) \\ \sqrt{f(1-f)}(1-p) & (1-p)(1-f)
\end{array} \right).
\end{equation}
The von Neumann entropy can be computed as $S(\rho)=-\lambda_+\log\lambda_+
-\lambda_-\log\lambda_-$, where $\lambda_\pm$ are the eigenvalues of $\rho$, i.e.
\begin{equation}
\lambda_\pm=\frac{1}{2}\pm \left(\frac{1}{4}-p(1-p)(1-f)\right)^{\frac{1}{2}}.
\end{equation}
We can see $dS/df<0$ for all $p$. Therefore, in order not to make the entropy decrease in time,
the change of $f$ must be non-positive: $|\braket{\phi(t)}{\psi(t)}|^2\le
|\braket{\phi(0)}{\psi(0)}|^2$.

Now let $\{\ket{\phi_k}\}$ be a complete orthogonal set spanning the whole Hilbert space. Then,
for any pure state $\ket{\psi}$, $\sum_k |\braket{\phi_k}{\psi}|^2=1$. Thus, if there is some $m$
for which $|\braket{\phi_m(t)}{\psi(t)}|^2<|\braket{\phi(0)}{\psi(0)}|^2$, there must be some $n$
for which $|\braket{\phi_n(t)}{\psi(t)}|^2>|\braket{\phi(0)}{\psi(0)}|^2$, which means that the
entropy of a mixture of $\ket{\phi_n}\bra{\phi_n}$ and $\ket{\psi}\bra{\psi}$ will
\textit{decrease} in a closed system. Hence, $f=|\braket{\phi}{\psi}|^2$ needs to be constant for
any $\ket{\phi}$ and $\ket{\psi}$ to comply with the second law. There are still two possibilities
for the time evolution of states $\ket{\psi(0)}\rightarrow \ket{\psi(t)}$ to keep $f$ constant,
namely unitary and antiunitary evolutions, according to Wigner's theorem \cite{wigner59}.
Nevertheless, the latter possibility can be excluded due to the continuity requirement. Therefore,
the evolution of quantum states is unitary, which is linear.

\subsection{Second law and general relativity}
\begin{figure}
\begin{center}
\includegraphics[scale=0.75]{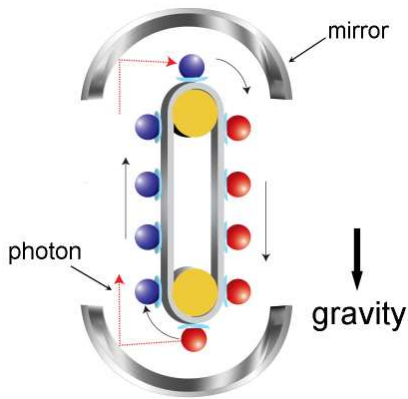}
\caption{Bondi's thought experiment for a perpetuum mobile. Excited atoms (red balls) emit a
photon at the bottom of the belt to lower its energy level (blue balls). The emitted photon is
reflected by the curved mirrors placed so that it will be absorbed by an atom in the lower energy
level at the top of the belt.} \label{bondi}
\end{center}
\end{figure}

The second law of thermodynamics gives an interesting implication not only in quantum mechanics,
but also in the theory of gravity, through the impossibility of the second kind of perpetuum
mobile. This is illustrated by Bondi's thought experiment. Although an assumption in the idea
presented below seems quite infeasible, let us have a look at it because it is a nice heuristic
introduction to discuss `real' physics later. Imagine a vertically placed conveyor belt that has a
number of single-atom holders on it as in Fig.~\ref{bondi}. Let us assume that the atoms on the
left side are in an excited state and those on the right side are in a lower energy state. When an
excited atom reaches the bottom of the belt, it emits a photon, lowering its energy level, and the
emitted photon will be reflected by the curved mirrors to be directed to the atom at the top of
the belt. Then this atom at the top will be excited, absorbing the photon.

As energy is equivalent to mass, according to special relativity, the atoms on the right side are
always heavier than those on the left as far as the emission and absorption of photon work as
described above. That is, the gravitational force will keep the belt rotating forever. In this
scenario, however, there is another assumption, which seems implausible, that atoms emit/absorb
photons only at the bottom/top of the belt. Such an assumption makes this device unlike to work.
Nevertheless, Bondi's perpetuum mobile is not compatible with the physical laws for the following
reason.

What prevents this machine from perpetual motion is actually the distorted spacetime, i.e. a
change of the metric, which is seen as gravity by us. The theory of general relativity tells that
the space is more stretched if one goes farther from the `horizon': the length of the geodesic
line is longer near the horizon for a given length in the normal sense, which is defined as the
circumference of the sphere around the massive object divided by $2\pi$. Because light travels
along a geodesic, stationary observers see that the wavelength becomes longer when light leaves
away from the object: the light becomes \textit{red-shifted}.

An experiment to confirm this red-shift was carried out in 1960 by Pound and Rebka \cite{pound60}.
They made use of the M\"{o}ssbauer effect of nuclear resonance, which can be used to detect
extremely small changes in frequency. Their results demonstrated that the photons do change the
frequency by a few parts of $10^{15}$ when they travel for 22.5m vertically, which agreed
Einstein's prediction with a high accuracy (only one percent error in the end \cite{pound64}). The
existence of the gravitational red-shift directly rules out Bondi's perpetuum mobile.

\subsection{Einstein equation from thermodynamics}
Einstein's equation, which describes the effect of energy-mass on the geometrical structure of the
four dimensional spacetime, can be derived from a fundamental thermodynamic relation. In
thermodynamics, knowing the entropy of a system as a function of energy and volume is enough to
get the equation of state from the fundamental relation, $\delta Q=TdS$. Jacobson tried to obtain
the field equation as an equation of state, starting from thermodynamic properties of black holes
\cite{jacobson95}. It had been known by then that there was a strong analogy between the laws of
black hole mechanics and thermodynamics \cite{bardeen73}. That is, the horizon area of a black
hole does not decrease with time, just as the entropy in thermodynamics. Bekenstein then argued
that the black hole entropy should be proportional to its horizon area after introducing the
entropy as a measure of information about the black hole interior which is inaccessible to an
exterior observer \cite{bekenstein73,bekenstein74}.

Bekenstein's idea suggests that it is natural to regard the (causal) horizon as a diathermic wall
that prevents an observer from obtaining information about the other side of it. On the other
hand, a uniformly accelerated observer sees a black body radiation of temperature $T$ from vacuum
(the Unruh effect) \cite{unruh76,davies75}. The origin of the Unruh effect lies in the quantum
fluctuation of the vacuum, which is also the origin of the entropy of the horizon, i.e. the
correlation between both sides of the horizon. Thus, in order to start from the above relation,
$\delta Q=TdS$, Jacobson associated $\delta Q$ and $T$ with the energy flow across the causal
horizon and the Unruh temperature seen by the observer inside the horizon, respectively. Then,
Einstein's field equation can be obtained by expressing the energy flow in terms of the
energy-momentum tensor $T_{\mu\nu}$ and the (horizon) area variation in terms of the `expansion'
of the horizon generators. Another essential element in the field equation, the Ricci tensor
$R_{\mu\nu}$, appears in the form of the expansion through the Raychaudhuri equation (for example,
see Ref.~\cite{poisson04}), which describes the rate of the volume (area) change of an object in a
Riemannian manifold. The resulting equation is thus (by setting $c=1$)
\begin{equation}
R_{\mu\nu}-\frac{1}{2}R g_{\mu\nu}+\Lambda g_{\mu\nu} = \frac{2\pi k}{\hbar\eta} T_{\mu\nu},
\end{equation}
where $\eta$ is the proportionality constant between the entropy and horizon area, viz. $dS=\eta
d\mathcal{A}$, and $R$ and $\Lambda$ are the scalar curvature and the cosmological constant.
Comparing with the standard expression of the equation, in which the coefficient for $T_{\mu\nu}$
is $8\pi G$, we identify $\eta$ to be $k/(4\hbar G)=k/(4l_P^2)$ with the Planck length $l_P$,
which agrees with the derivation of $\eta$ in Ref.~\cite{hawking74}.

Einstein's field equation can indeed be seen as a thermodynamic equation of state. An important
assumption for the above derivation is, however, the existence of a local equilibrium condition,
for which the relation $\delta Q=TdS$ is valid. This means that it would not be appropriate to
quantize the field equation as it is not appropriate to quantize the wave equation for sound
propagation in air. Further, Jacobson speculated that the Einstein equation might not describe the
gravitational field with sufficiently high frequency or large amplitude disturbances, because the
local equilibrium conditions would break down in such situations as in the case of sound waves.

Bekenstein's conjecture about the black hole entropy is now widely accepted as a real physical
property, particularly after the discovery of the Hawking radiation \cite{hawking74} that showed
that black holes do radiate particles in a thermal distribution at finite temperature. The
thermodynamics of black holes is still an extensive and active field, whose famous problems
include the `black hole information paradox'. Covering these topics in detail go beyond the scope
of this brief overview, thus here we simply list a few references
\cite{preskill92,jacobson_lectnotes,wald01,bousso02,hawking05,page05}.

It is now clear that this example illustrates a close connection between information,
thermodynamics, and the general relativity, which might look unrelated with each other at first
sight. This strongly re-suggests the duality of entropy, which we have mentioned at the end of
Section \ref{temp_solution}, and the universality of the thermodynamic relations in generic
physics. We shall attempt to explore this duality in the paradigm of quantum information theory
later on in Section \ref{holevo_from_2ndlaw}.

\section{Erasure of classical information encoded in quantum states}\label{lubkin_erasure}
In classical information theory, an alphabet $i\in\{1,...,n\}$, which appears with probability
$p_i$ in a message \footnote{A message can be any set of alphabets. It refers to a word/letter
sent from sender to receiver, information which is stored in memory, etc. We assume that the
information source generates independent and identically distributed variables/alphabets according
to the probability distribution $\{p_i\}$.} generated by a source, is represented by one of the
$n$ different classical states. On the other hand, in quantum information theory
\cite{nielsenchuang}, information is encoded in quantum states, so that each alphabet $i$ is
represented by one of the $n$ different quantum states whose density matrices are denoted by
$\rho_i$. We will refer to each state carrying an alphabet as a message state. We will also call a
set of quantum states used in a message, in which the state $\rho_i$ appears with probability
$p_i$, an ensemble of quantum states $\{p_i, \rho_i\}, i\in\{1,...,n\}$.

A smart way to erase classical information encoded in quantum states was first considered by
Lubkin \cite{lubkin87}, who introduced erasure by thermal randomization, and by Vedral
\cite{vedral00,vedral02} in a more general setting. Thermal randomization makes use of the
randomness of states in a heat bath that is in thermal equilibrium. If we put a message state in
contact with a heat bath at temperature $T$, the state will approach thermal equilibrium with the
heat bath. More precisely, a message state $\rho_i$ changes gradually after colliding
(interacting) with the heat bath and sufficiently many collisions make the state to become
indistinguishable with that of the heat bath. We assume that the bath's state as a whole will not
change much since its size is very large. Due to the uncertainty stemming from thermal
fluctuations, we irreversibly lose the information that was carried by the state $\rho_i$.

\begin{figure}
\begin{center}
\includegraphics[scale=0.5]{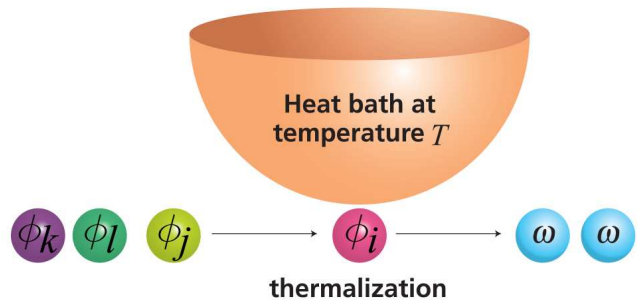}
\caption{The erasure of classical information carried by quantum states. Each message state
interacts with a heat bath at temperature $T$ and reaches thermal equilibrium. The information
originally encoded in a state is lost and all states end up in $\omega$, which is the thermal
state at the temperature $T$.} \label{lubkin}
\end{center}
\end{figure}

Because of the generic nature of this erasure process by thermalization, entropy of the whole
system, consisting of the message state and the heat bath, necessarily increases. How much would
this increase be? Let us first simplify the discussion by considering that each message state is a
pure state as in Fig.~\ref{lubkin} \cite{vedral00,plenio01b}. Before erasing, the whole message is
an ensemble $\{p_i,\ket{\phi_i}\}$, thus its average state is described by a density operator
$\rho=\sum_i p_i \ket{\phi_i}\bra{\phi_i}$. The thermalization process brings all states
$\ket{\phi_i}$ to the same state $\omega$, which is in thermal equilibrium at temperature $T$. The
density matrix $\omega$ is given by
\begin{equation}\label{gibbs}
\omega=\frac{e^{-\beta H}}{Z}=\sum_j q_j\,\ket{e_j}\bra{e_j},
\end{equation}
where $H=\sum_i e_i\,\ket{e_i}\bra{e_i}$ is the Hamiltonian of the message state with energy
eigenstates $\ket{e_i}$, $Z=\tr(e^{-\beta H})$ is the partition function, and $\beta=(kT)^{-1}$.

The total entropy change $\Delta S_\mathrm{erasure}$ is the sum of the entropy change of the
message system and that of the heat bath: $\Delta S_\mathrm{erasure}=\Delta S_\mathrm{sys}+\Delta
S_\mathrm{bath}$. Since the message state before the erasure is pure and its state after the
erasure is the same as the heat bath, the minimum entropy change in the message state is given by
\footnote{One may be tempted to use the averaged message state $\rho$ as the pre-erasure state.
However, this is not the right way of viewing it. Before the erasing procedure, the encoder, who
prepared the state, or the memory itself still knows which of $\{\ket{\psi_i}\}$ it is in.
Information erasure is a process that destroys correlations between the memory and the encoder or
the system accessing to it, by transforming the state to a standard state (i.e.
$\omega\simeq\ket{e_0}\bra{e_0}$ in Lubkin's erasure), irrespective of the initial state. In other
words, there must be a perfect correlation or knowledge before the erasure, which will be lost
afterwards. Averaging over an ensemble means that even the encoder already lost information about
his/her preparation. Hence, in this case, the entropy of the pre-erasure state should be taken as
0. Considering the classical counterpart (Fig.~\ref{erasure_gas}(a)) may be useful to understand
this reasoning.}
\begin{equation}\label{deltas_sys}
\Delta S_\mathrm{sys}=k\ln2 S(\omega),
\end{equation}
where $S(\omega)=-\tr(\omega\log\omega)$ is the von Neumann entropy of the state
$\omega$. Von Neumann introduced this entropy by contemplating the disorder of quantum
states so that it has the same meaning as the entropy in (phenomenological)
thermodynamics in a setting where gases of molecules with quantum properties were
considered \cite{vonneumann}. The factor $k\ln2$ is just a conversion factor to make it
consistent with the previous discussion of Landauer's principle.

The entropy change in the heat bath is equal to the average heat transfer from the bath to the
message system divided by the temperature: $\Delta S_\mathrm{bath}=\Delta Q_\mathrm{bath}/T$. The
heat change in the heat bath is the same as that in the system with an opposite sign, i.e. $\Delta
Q_\mathrm{bath}=-\Delta Q_\mathrm{sys}$. The heat transfer can be done quasistatically so that the
mechanical work required for the state change is arbitrarily close to 0. Therefore, due to the
energy conservation, $\Delta Q_\mathrm{sys}$ must be equal to the change of internal energy of the
message system $\Delta U_\mathrm{sys}$, which can be computed as the change of average values of
the Hamiltonian $H$ before and after the erasure process. Hence,
\begin{eqnarray}\label{deltas_bath1}
\Delta S_\mathrm{bath} &=& -\frac{\Delta Q_\mathrm{sys}}{T}=-\frac{\Delta U_\mathrm{sys}}{T} \nonumber \\
&=& -\frac{\tr(\omega H)-\tr(\rho H)}{T} \nonumber \\
&=& -\frac{\tr\left[(\omega-\rho)H\right]}{T}.
\end{eqnarray}
By using Eq.~(\ref{gibbs}), the Hamiltonian $H$ can be expressed in terms of the partition
function $Z$ as $-kT\ln(Z\omega)$. Now we have
\begin{eqnarray}\label{deltas_bath2}
\Delta S_\mathrm{bath} &=& k\tr\left[(\omega-\rho)\ln(Z\omega)\right] \nonumber \\
&=& k\tr\left[(\omega-\rho)\ln\omega\right] \nonumber \\
&=& -k\ln2 [S(\omega)+\tr(\rho\log\omega)].
\end{eqnarray}
Combining Eq.~(\ref{deltas_sys}) and Eq.~(\ref{deltas_bath2}) gives the total entropy change after
the erasure:
\begin{equation}\label{deltas_erasure}
\Delta S_\mathrm{erasure}=\Delta S_\mathrm{sys}+\Delta S_\mathrm{bath}=-\tr(\rho\log\omega),
\end{equation}
where the unimportant conversion factor $k\ln2$ is set to be unity as a unit of entropy. The
minimum of the entropy change $\Delta S_\mathrm{erasure}$ can be obtained as:
\begin{equation}\label{deltas_minimum}
\Delta S_\mathrm{erasure}=-\tr(\rho\log\omega)\ge S(\rho).
\end{equation}
The inequality follows from the property of the quantum relative entropy,
$S(\rho||\omega):=-S(\rho)-\tr (\rho\log\omega)\ge 0$. This minimum can be achieved by choosing
the temperature of the heat bath and the set $\{p_i,\ket{\phi_i}\}$ such that $\rho=\sum
p_i\ket{\phi_i}\bra{\phi_i}$ is the same as the thermal equilibrium state $\omega$. Consequently,
the minimum entropy increase required for the erasure of classical information encoded in quantum
states is given by the von Neumann entropy $S(\rho)$, where $\rho$ is the average state of the
system, instead of the Shannon entropy $H(p)$ in the case of erasing information in classical
states.

\section{Thermodynamic derivation of the Holevo bound}\label{holevo_from_erasure}
\subsection{from the erasure principle}
Landauer's erasure principle, together with its Lubkin's version for quantum states, is simple in
form; however, it implies some significant results in the theory of quantum information. For
example, it can be used to derive the efficiency of the compression of data carried by quantum
states \cite{plenio01b} and also the upper bound on the efficiency of the entanglement
distillation process \cite{vedral98,vedral00}. Here, we look at the derivation of the Holevo bound
from Landauer's principle, which was first discussed by Plenio \cite{plenio99,plenio01b}, as we
will examine the same problem from a different perspective in the next section.

To give a precise form of the Holevo bound let us consider two parties, Alice and Bob. Suppose
Alice has a classical information source preparing symbols $i=1,...,n$ with probabilities
$p_1,...,p_n$. The aim for Bob is to determine the actual preparation $i$ as best as he can. To
achieve this goal, Alice prepares a state $\rho_i$ with probability $p_i$ and gives the state to
Bob, who makes a general quantum measurement (\textit{Positive Operator Valued Measure} or
\textit{POVM}) with elements ${E_j}={E_1,...,E_m}$, $\sum_{j=1}^{m} E_j =\mathbf{l}$, on that
state. On the basis of the measurement result he makes the best guess of Alice's preparation. The
Holevo bound \cite{holevo73} is an upper bound on the accessible information, i.e.
\begin{equation}\label{holevo00}
I(A:B)\le S(\rho)-\sum_i p_i S(\rho_i),
\end{equation}
where $I(A:B)$ is the mutual information between the set of Alice's preparations $i$ and Bob's
measurement outcomes $j$, and $\rho=\sum_i^n p_i \rho_i$. The equality is achieved when all
density matrices commute, namely $[\rho_i,\rho_j]=0$.

Let us first consider a simple case in which all $\rho_i$ are pure:
$\rho_i=\ket{\psi_i}\bra{\psi_i}$. The average state will be $\rho=\sum_i p_i
\ket{\psi_i}\bra{\psi_i}$. Then, the Shannon entropy of the message is always greater than or
equal to the von Neumann entropy of the encoded quantum state (Theorem 11.10 in
Ref.~\cite{nielsenchuang}), that is, $H(p_i)\ge S(\rho)$, with equality if and only if
$\braket{\psi_i}{\psi_j}=\delta_{ij}$ for all $i$ and $j$.

How much information can Bob retrieve from the state $\rho$? The above analysis of erasure of
information in quantum states tells us that we have to invest at least $S(\rho)$ bits of entropy
to destroy all available information. This implies that the amount of information that Bob can
access is bounded by $S(\rho)$, because if he could obtain $S(\rho)+\varepsilon$ bits of
information then the minimum entropy increase by the erasure should be at least
$S(\rho)+\varepsilon$, by Landauer's principle. In other words, one cannot obtain more information
than it could be erasable.

Therefore, the accessible information $I(A:B)$ is smaller than or equal to the minimum entropy
increase by the erasure and it is this relation that corresponds to the inequality
(\ref{holevo00}). If Alice encodes her information $i$ into a pure state $\rho_i$, the relation
reads
\begin{equation}\label{erasure_holevo_pure}
I(A:B)\le S(\rho),
\end{equation}
which is the same inequality as Eq.~(\ref{holevo00}) because $S(\rho_i)=0$ for all pure states
$\rho_i$.

If Alice uses mixed states $\rho_i$ to encode $i$, Eq.~(\ref{deltas_minimum}) needs to be
modified. Instead of Eq.~(\ref{deltas_sys}), the entropy increase for each state is now
\footnote{The conversion factor $k\ln2$ is set to be equal to unity again as a unit of entropy.}
$\Delta S_\mathrm{sys}^j=S(\omega)-S(\rho_j)$ after contact with a heat bath whose state is given
by Eq.~(\ref{gibbs}). The average entropy change of the heat bath is the same as
Eq.~(\ref{deltas_bath2}): $\Delta S_\mathrm{bath}=-S(\omega)-\rho\log\omega$. The total entropy
change by the thermalization will be
\begin{eqnarray}\label{deltas_erasure_mixed}
\Delta S_\mathrm{erasure} &=& \sum_j p_j\Delta S_\mathrm{sys}^j+\Delta S_\mathrm{bath}
\nonumber \\
&=&  \sum_j p_j(S(\omega)-S(\rho_j))-S(\omega)-\tr(\rho\log\omega) \nonumber \\
&=&  -\sum_j p_j S(\rho_j)-\tr(\rho\log\omega) \nonumber \\
&\ge&  S(\rho)-\sum_j p_j S(\rho_j),
\end{eqnarray}
which, together with the above argument, implies the Holevo bound in the form of
Eq.~(\ref{holevo00}). This analysis of erasure of information encoded with mixed states is more
straightforward and less ambiguous than that in Refs.~\cite{plenio99,plenio01b}.

The above analysis thus justifies the Holevo bound. However, it does not give the precise
condition for the equality in Eq.~(\ref{holevo00}), which is $[\rho_i,\rho_j]=0$. The condition we
can derive here is that all density matrices $\{\rho_i\}$ support orthogonal subspaces, i.e.
$\tr(\rho_i\rho_j)=0$. This is more restrictive than the commutativity of density matrices
mentioned above. In the next section, we will see if the second law implies the Holevo bound more
directly.

\subsection{from the second law}\label{holevo_from_2ndlaw}
Since the work-information \textit{duality} in the erasure principle supports Brillouin's
hypothesis, which we have seen in Section \ref{temp_solution}, on the equivalence between
information theoretic and thermodynamic entropies, it might be natural to expect that the second
law may put a certain bound on the quality of information or the performance of information
processing. In this section, we derive the general bound on storage of quantum information, the
Holevo bound \cite{holevo73} derived from the second law of thermodynamics. As the second law is
the most fundamental physical law that governs the behavior of entropy, this problem is
interesting in terms of the spirit of the `physics of information' and deserves to be investigated
in its own right.

In order to see the genuine thermodynamic bound, we need to minimize the axiomatic assumptions
that stem from quantum mechanics. Assumptions we make here are, (a) Entropy: the von Neumann
entropy is equivalent to the thermodynamic entropy, (b) Statics and measurement: a physical state
is described by a ``density" matrix, and the state after a measurement is a new state that
corresponds to the outcome (``projection postulate"), (c) Dynamics: there exist isentropic
transformations.

Employing the density matrix-based description means that we presume the existence of
superpositions of states. Allowing superpositions might sound rather abrupt; however, we can
assume that we are taking a similar stand as Gabor's picture on the possibility of superpositions
purely from thermodynamic considerations. Thus, this could be stated in the other way around:
assuming the superpositions of states, we can describe a state by a density matrix, which can be
defined as a convex sum of outer products of normalized ``state vectors". The nonzero components
of state vectors represent superposed state elements. Probability distributions in classical phase
space can also be described consistently: all diagonal elements of a classical density matrix are
real, representing probabilities, and all off-diagonal elements are zero. When a measurement is
performed, one of the diagonal elements becomes 1, replacing all others with 0. We will use an
arrow to denote a state vector, such as $\vec{\psi}$, to make it clear that we do not use the full
machinery of the Hilbert space (such as the notion of inner product) and we never use the Born
trace rule for calculating probabilities.

Consider a chamber of volume $V$, which is divided into two regions of volumes $p_1 V$ and $p_2
V$, respectively. The left-side region ($L$) is filled with $p_1 N$ molecules in state
$\vec{\psi}_1$, and $p_2 N$ molecules in state $\vec{\psi}_2$ are located in the right-side region
($R$). The two states, $\vec{\psi}_1$ and $\vec{\psi}_2$, can be thought of a representation of an
internal degree of freedom. Generalizations to arbitrary numbers of general (mixed) states and
general measurements are straightforward.

\begin{figure}
 \begin{center}
  \includegraphics[scale=0.5]{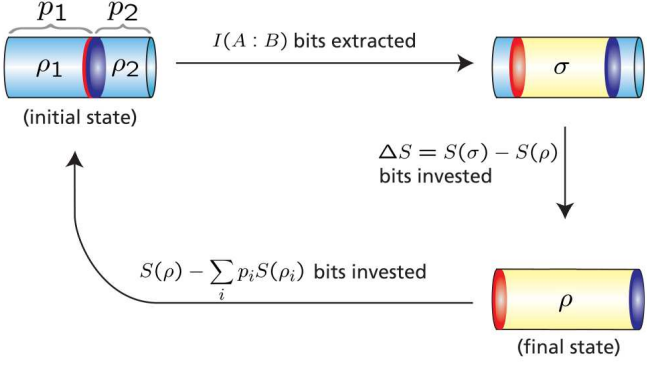}
  \caption{The thermodynamic cycle, which we use to discuss the second law. The cycle
proceeds from the initial state (a) to the final state $\rho$ (c) via the post-work-extraction
state $\sigma$ (b), and returns to the initial state with a reversible process.}
 \label{cycle}
 \end{center}
\end{figure}

We can now have a thermodynamic loop formed by two different paths between the above initial
thermodynamic state to the final state (Fig.~\ref{cycle}). In the final state, both constituents,
$\vec{\psi}_1$ and $\vec{\psi}_2$, are distributed uniformly over the whole volume. Hence, each
molecule in the final state can be described by $\rho=\sum_i p_i
\vec{\psi}_i\vec{\psi}_i^\dagger$, regardless of the position in the chamber. One of the paths
converts heat into work, while the other path, consisting of a quasi-static reversible process and
isentropic transformations, requires some work consumption.

In the work-extracting process, we make use of two semipermeable membranes, $M_1$ and $M_2$, which
separate two perfectly distinguishable (orthogonal) states $\vec{e}_1$ and
$\vec{e}_2\;(=\vec{e}_1^\perp)$. The membrane $M_i$ ($i=1,2$) acts as a completely opaque wall to
molecules in $\vec{e}_i$, but it is transparent to molecules in $\vec{e}_{j\neq i}$. Thus, for
example, a state $\vec{\psi}_i$ is reflected by $M_1$ to become $\vec{e}_1$ with (conditional)
probability $p(e_1|\psi_i)$ and goes through with probability $p(e_2|\psi_i)$, being projected
onto $\vec{e}_2$. This corresponds to the quantum (projective) measurement on molecules in the
basis $\{\vec{e}_1,\vec{e}_2\}$, however, we do not compute these probabilities specifically as
stated above.

\begin{figure}
 \begin{center}
  \includegraphics[scale=0.5]{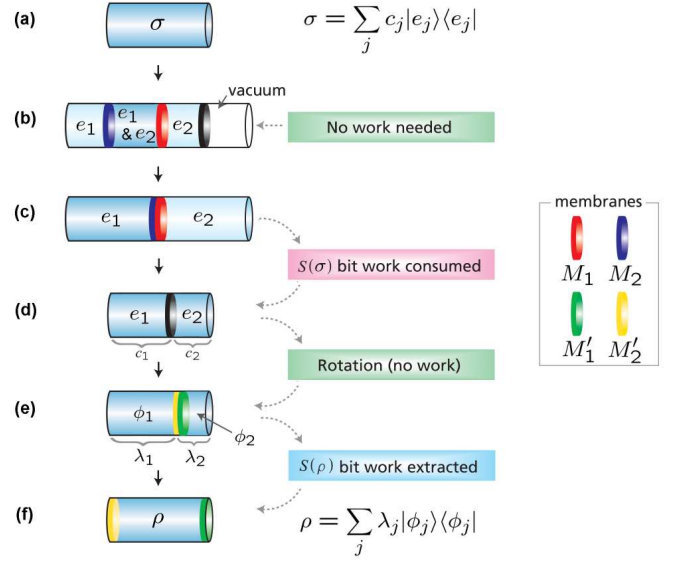}
  \caption{The thermodynamic process to transform the intermediate state $\sigma$ into the final state $\rho$.
Firstly, after attaching an empty vessel of the same volume to that containing the gas $\sigma$,
the membranes $M_j$ are used to separate two orthogonal states $\vec{e}_1$ and $\vec{e}_2$ ((a) to
(c)). As the distance between the movable opaque wall and the membrane $M_2$ is kept constant,
this process entails no work consumption/extraction. As $\sigma=\sum
c_j\vec{e}_j\vec{e}_j^\dagger$, compressing each $\vec{e}_j$-gas into the volume of $c_jV$ as in
(d) makes the pressures of gases equal and this compression requires $S(\sigma)=-\sum c_j\log_2
c_j$ bits of work. Second, quantum states of gases are isentropically transformed, thus without
consuming work, so that the resulting state (e) will have $\lambda_j N$ molecules in
$\vec{\phi}_j$, where $\rho=\sum \lambda_j\vec{\phi}_j\vec{\phi}_j^\dagger$ is the
eigendecomposition of $\rho$. To reach (f), $S(\rho)$ bits of work can be extracted by using
membranes that distinguish $\vec{\phi}_j$. As a result, the work needed for the transformation
$\sigma\rightarrow\rho$ is $S(\sigma)-S(\rho)$ bits.}
 \label{sigma2rho}
 \end{center}
\end{figure}

By replacing the impenetrable partition with the two membranes, we can convert heat from the heat
bath into mechanical work $W_\mathrm{ext}$, which can be as large as the accessible information
$I(A:B)$, i.e. the amount of information Bob can obtain about Alice's preparation by measurement
in the basis $\{e_1, e_2\}$ \footnote{The justification of this equivalence described in Fig.~1 of
Ref.~\cite{maruyama05b} has an error in identifying entropy changes in the process. Nevertheless,
this equivalence between $W_\mathrm{ext}$ and $I(A:B)$ can be seen correct by a bit of
straightforward calculations, using the state equation for an ideal gas.}. The transformation from
the post-work-extraction state, which we call $\sigma$ hereafter, to the final state $\rho$ can be
done by a process shown in Fig.~\ref{sigma2rho} and the minimum work needed is given by $\Delta
S=S(\sigma)-S(\rho)$.

Another path, which is reversible, from the initial state to the final state is as follows. Let
$\{\vec{\phi}_1,\vec{\phi}_2\}$ be an orthonormal basis which diagonalizes the density matrix
$\rho$, such that
\begin{equation}\label{decompositions}
\rho=\sum_i p_i \vec{\psi}_i\vec{\psi}_i^\dagger=\sum_k \lambda_k
\vec{\phi}_k\vec{\phi}_k^\dagger,
\end{equation}
where $\lambda_k$ are eigenvalues of $\rho$. We can extract $S(\rho)$ bits of work by first
transforming $\{\vec{\psi}_1,\vec{\psi}_2\}$ to $\{\vec{\phi}_1,\vec{\phi}_2\}$ unitarily, and
second using a new set of semipermeable membranes that perfectly distinguish $\vec{\phi}_1$ and
$\vec{\phi}_2$.

If the initial state is a combination of mixed states with corresponding weights given by
$\{p_i,\rho_i\}$, the extractable work during the transformation to $\rho=\sum_i p_i\rho_i$
becomes $S(\rho)-\sum_i p_i S(\rho_i)$. This can be seen by considering a process
\[
\{p_i,\rho_i\}\stackrel{\mbox{\small{(i)}}}{\longrightarrow}
\{p_i\mu_j^i,\vec{\omega}_j^i\}\stackrel{\mbox{\small{(ii)}}}{\longrightarrow}
\{\lambda_k,\vec{\phi}_k\}\stackrel{\mbox{\small{(iii)}}}{\longrightarrow}\rho,
\]
where $\{\mu_j^i,\vec{\omega}_j^i\}$ and $\{\lambda_k,\vec{\phi}_k\}$ are the sets of eigenvalues
and eigenvectors of $\rho_i$ and $\rho$, respectively \cite{maruyama05b}.

The function of the semipermeable membrane can alternatively be understood as a Maxwell's demon
who controls small doors on a partition depending on the result of his measurement of each
molecule. Then do we need to consume some work to reset his memory? Unlike the previous
discussions (such as that of Szilard's engine), it turns out that the demon's memory can be erased
isentropically due to the remaining (perfect) correlation between the state of each molecule and
his memory registers. This can be sketched as follows. Once the demon observes a molecule, a
correlation between the state of the molecule and his memory is created. Since he can in principle
keep track of all molecules, a perfect correlation between the state of the $n$-th molecule and
that of the $n$-th register of demon's memory will be maintained.  Then a controlled-NOT-like
isentropic operation between the molecules and the corresponding memory registers (with molecules
as a control \textit{bit}) can reset the demon's memory to a standard initial state without
consuming work.

The second law states (in Kelvin's form) that the \textit{net} work extractable from a heat bath
cannot be positive after completing a cycle, i.e. $W_\mathrm{ext}-W_\mathrm{inv}\le 0$. For the
cycle described in Fig.~\ref{cycle}, it can be expressed as
\begin{equation}\label{seclaw}
I(A:B)\le S(\rho)-\sum_i p_i S(\rho_i)+\Delta S,
\end{equation}
where $\Delta S=S(\sigma)-S(\rho)$. As $\sigma$ is identical to the resulting state of a
projective measurement on $\rho$ in the basis $\{\vec{e}_1,\vec{e}_2\}$, $\sigma=\sum_j P_j\rho
P_j$ with $P_j=\vec{e}_j\vec{e}_j^{\,\dagger}$ and consequently $\Delta S$ is always non-negative
(See Ref.~\cite{vonneumann}). The inequality (\ref{seclaw}) holds even if the measurement by
membranes was a generalized (POVM) measurement \cite{maruyama05b}.

The form of Eq.~(\ref{seclaw}) is identical with that of Eq.~(\ref{holevo00}) for the Holevo
bound, except an extra non-negative term, $\Delta S$. This illustrates that there is a difference
between the bound imposed by quantum mechanics (the Holevo bound) and the one imposed by the
second law of thermodynamics. Namely, there is a region in which we could violate quantum
mechanics while complying with the thermodynamical law. In the classical limit, the measurement is
performed in the joint eigenbasis of mutually commuting $\rho_i$'s, consequently $\Delta S=0$,
and, in addition, the Holevo bound is saturated: $I(A:B)= S(\rho)-\sum_i p_i S(\rho_i)$. Thus, the
classical limit and the thermodynamic treatment give the same bound.

The same saturation occurs when an appropriate collective measurement is performed on blocks of
$m$ molecules, each of which is taken from an ensemble $\{p_i,\rho_i\}$. When $m$ tends to
infinity $2^{m(S(\rho)-\sum_i p_i S(\rho_i))}$ typical sequences (the sequences in which $\rho_i$
appears about $p_i m$ times) become mutually orthogonal and can be distinguished by ``square-root"
or ``pretty good" measurements \cite{hausladen96,holevo98}. This situation is thus essentially
classical, hence, $\Delta S\rightarrow 0$ and the Holevo bound will be saturated.

\section{Entanglement detection by Maxwell's demon(s)}
Now let us move on to see how we can deal with entanglement from the point of view of Maxwell's
demon. The reason we pick up the topic of entanglement in particular is that it is not only
crucially important in quantum information theoretic tasks, such as quantum cryptography
\cite{bennett84,ekert91}, dense coding \cite{bennett92}, quantum teleportation \cite{bennett93}
and quantum computation \cite{shor94,grover96}, but it is also directly linked to the foundations
of quantum mechanics. However, the theory of entanglement is too broad and deep to explore
comprehensively in this article. In addition, the most of it appears irrelevant or at least
unclear in the context of Maxwell's demon at any rate. Therefore we focus on only some recent work
that have discussed the `quantumness' of correlation and/or the problem of entanglement detection,
which is one of the most important topics in its own right.

When we discuss entanglement in this article we are primarily interested in bipartite
entanglement, unless otherwise stated. Formally, entanglement is defined as a form of quantum
correlation that is not present in any \textit{separable} states. Let $\mc{H}^A$ and $\mc{H}^B$ be
the Hilbert spaces for two spatially separated (non-interacting) subsystems $A$ and $B$, which are
typically referred to as Alice and Bob, and $\mc{H}^{A\!B}=\mc{H}^A\otimes \mc{H}^B$ the whole
(joint) Hilbert space of the two. We also let $\mc{S}(\mc{H})$ denote the \textit{state space},
which is a set of density operators acting on $\mc{H}$.

A state of a bipartite system is said to be separable or classically correlated \cite{werner89} if
its density operator can be written as a convex sum of products of density operators
\begin{equation}\label{separability00}
\rho=\sum_{i=1}^{n}p_i \rho_i^A \otimes \rho_i^B,
\end{equation}
where all $p_i$ are nonnegative and $\sum_{i=1}^n p_i=1$. Any state that cannot be written in the
form of Eq.~(\ref{separability00}) is called \textit{entangled}. We will let $\mc{S}_\mathrm{sep}$
denote the subspace that contains all separable states.

It is natural to ask whether or not a given state $\rho\in \mc{S}(\mc{H}^{A\!B})$ is separable,
considering the importance of entanglement in many quantum information theoretic tasks
\footnote{Looking at entanglement as a resource of quantum information processing naturally
suggests the way to quantify entanglement in terms of its usefulness for such tasks. This leads to
the idea of \textit{distillable entanglement}, i.e. the average number of maximally entangled
pairs that can be distilled from a given pair using only local operations and classical
communication (LOCC) \cite{bennett96b,rains01}.}\label{fn_distillableent}. Quite a few
\textit{separability criteria}, i.e. a condition that is satisfied by all separable states, but
not necessarily by entangled states, have been proposed so far to answer this simple question.
Separability criteria are typically expressed in terms of an operator or a function, such as an
entanglement witness \cite{terhal00} or the correlation function in Bell's inequality
\cite{bell64}. Despite the simplicity of the question, it is generally very hard to find good
separability criteria. By a \textit{good} separability criterion, we mean an efficient
separability criterion that singles out as many entangled states as possible. The hardness of the
problem is primarily related to the convexity of the separable subspace, which is formed by all
the separable states: Because of the bulgy `surface' of the separable subspace, there does not
exist any operator/function that is linear with respect to the matrix elements of density
operator, e.g. eigenvalues of $\rho$, and distinguishes separable and entangled states perfectly.

Another simple question is about the \textit{amount} of entanglement a pair (or a set) of quantum
objects contains. It plays a major role when it comes to the characterization or manipulation of
entanglement. Since entanglement can be regarded as a valuable resource in quantum information
processing, the quantification of entanglement is a problem of great interest and importance.
Despite its profoundness, we will not go into details on the quantification issue here: Instead,
we refer interested readers to some references, such as
Refs.~\cite{Vedral97,schumacher00,horodecki01b,horodecki01c,plenio05,horodecki07}. Also,
Ref.~\cite{vedral02} contains not only a review on entanglement measures, but also some
discussions on quantum information processing from the thermodynamic point of view.

\subsection{Work deficit}\label{sec_deficit}
In this subsection, we will review the concept of \textit{work deficit}, which was introduced by
Oppenheim et al.~\cite{oppenheim02}. An apparent goal of this work was to quantify entanglement
via a thermodynamic quantity; this idea shed a new light on the quantumness of correlations by
taking a thermodynamic approach.

As we have emphasized, information is always stored in a physical system with physical states that
are distinguishable by measurement so that stored information can be extracted. No generality is
lost when we think of a gas in a chamber, such as the one considered by Szilard, as a general
information-storage apparatus. Even if we had a different type of physical system for information
storage, the information can be perfectly transferred for free to the memory of the type of
Szilard's engine, if the initial state of Szilard's engine is provided in a standard state. Since
the measurement can be done with negligible energy consumption (as we have seen in Section
\ref{exorcism}), the information transfer can be completed by converting the state from the
initial standard state to the state corresponding to the stored (measured) information; the final
conversion requires no energy.

Now that we have identified a memory with the one molecule gas of Szilard's engine we can present
a general statement: from an ensemble of memories, each of which stores the value of an $n$-bit
random variable $X$, one can extract mechanical work whose average amount per single memory
register is (by taking units such that $k\ln2=1$)
\begin{equation}\label{w_c}
W_C=n-H(X),
\end{equation}
where $H(X)$ is the Shannon entropy of $X$. The extractable work is the work done by the gas for
memory, thus it is nothing but Eq.~(\ref{exp-comp}) when $n=2$. Equation (\ref{w_c}) can be easily
understood in the following way. Suppose there are $N$ memory registers. If we measure all $N$
registers the remaining uncertainty in the memory is zero; we can obtain $Nn$ bits of work.
Nevertheless, we still keep the information due to the measurement on memory and this needs to be
erased to discuss solely the amount of extractable work. The minimum energy consumption to erase
the information is, according to the erasure principle, equal to $NH(X)$ bits. Thus the maximum
total amount of extractable work is given by $N(n-H(X))$ bits. Alternatively, one can use the
first law of thermodynamics to arrive at the same expression as Eq.~(\ref{w_c}). The work done by
the gas in an isothermal process is equal to the entropy change multiplied by the temperature.

The same argument is applicable to work extraction from quantum bits (qubits). Let $\rho$ be the
density operator describing the state in a given ensemble. Qubits after (non-collective)
measurements are in a known pure state, which is essentially a classical system in terms of
information. Thus the information stored in this set of pure states can be copied to the
Szilard-type memory and each register can give us 1 bit of work. Then, after erasing the
information acquired by the measurement, the net maximum amount of work we get becomes $1-S(\rho)$
bits of work.

The work deficit is a difference between the globally and the locally extractable work within the
framework of LOCC, i.e. local operations and classical communication, when $\rho$ is a system with
spatially separated subsystems. Suppose that we have an $n$ qubit state $\rho_{A\!B}$, which is
shared by Alice and Bob, then the optimal work extractable is
\begin{equation}\label{globalwork}
W_{\mr{global}}=n-S(\rho^{A\!B}),
\end{equation}
if one can access the entire system globally. On the other hand, we let $W_{\mr{local}}$ be the
largest amount of work that Alice and Bob can locally extract from the same system under LOCC. The
deficit $\Delta$ is defined as
\begin{equation}\label{deficit}
\Delta=W_{\mr{global}}-W_{\mr{local}}
\end{equation}

In order to grasp this picture, let us compute the deficits for a classically correlated state
\begin{equation}\label{clcorr}
\rho^{AB}_\mathrm{cl}=\frac{1}{2}(\ket{00}\bra{00}+\ket{11}\bra{11})
\end{equation}
and a maximally entangled state
\begin{equation}\label{maxent}
\ket{\Phi^{AB}}=\frac{1}{\sqrt{2}}(\ket{00}+\ket{11}).
\end{equation}
The globally extractable work $W_{\mr{global}}^\mathrm{cl}$ from $\rho^{AB}_\mathrm{cl}$ is simply
1 bit. The locally extractable work $W_{\mr{local}}^\mathrm{cl}$ turns out to be also 1 bit. The
protocol is as follows. Alice can measure her bit in the $\{\ket{0},\ket{1}\}$ basis and send the
result to Bob, who can obtain 1 bit of work from his bit. Although Alice can extract 1 bit of work
from her own bit, using her measurement result, she needs to consume this energy to erase the
information stored in the memory, which was used to communicate with Bob. Thus, the deficit for
the state $\rho^{AB}_\mathrm{cl}$ is $\Delta_\mathrm{cl}=1-1=0$. The locally extractable work is
the same, i.e. 1 bit, even if the state is maximally entangled as in Eq.~(\ref{maxent}). However,
as this state is pure globally, we can have
$W_{\mathrm{global}}=2-S(\ket{\Phi^{AB}}\bra{\Phi^{AB}})=2$ bits, therefore
$\Delta_\mathrm{ent}=2-1=1$.

These two simple examples suggest that the `strength' of correlation could be reflected in the
deficit, though the deficit might not be necessarily the amount of entanglement. In fact, the
authors of \cite{oppenheim02} propose later in a more detailed paper \cite{horodecki05} that the
(quantum) deficit can be interpreted as the amount of quantumness of correlations, not
entanglement.

It has been shown in  \cite{oppenheim02} that the deficit is bounded from below as $\Delta\ge
\mr{max}\{S(\rho^A),S(\rho^B)\}-S(\rho)$ (under an assumption about the classicality of the
communication channel), where $\rho^A$ and $\rho^B$ are the reduced density operators, i.e.
$\rho^A=\tr_B\rho$ and $\rho^B=\tr_A\rho$. The bound (or the upper bound for $W_{\mr{local}}$) can
be achieved when the state is pure and it turns out to be equal to the entanglement measure for
pure states. This is simply because a pure state can be written as $\ket{\psi}=\sum_i \alpha_i
\ket{e_i}\ket{f_i}$ in the Schmidt decomposition and then $\Delta=S(\rho^A)=E(\psi)$, where
$\rho^A=\tr_B \ket{\psi}\bra{\psi}$ and $E(\cdot)$ is the entanglement measure for pure states.

A similar approach has been taken in an attempt to quantify the amount of entanglement in
Ref.~\cite{groisman05}. There, the (asymptotically) minimal amount of noise added to the system to
erase the correlation was examined: roughly speaking, it can be characterized by the number of
allowed operations from which we choose randomly to make the given state separable. In the
discussion above on deficit, the correlation is converted into work and the purity of the system
is destroyed. Instead, noise is added actively here, and the information about the chosen
operations is erased in the end, dissipating entropy into the environment.

\subsection{Quantum discord}\label{sec_discord}
A similar approach to measuring the quantumness of correlations has been taken by Zurek
\cite{zurek03a} by using the concept of ``quantum discord" \cite{zurek00}, which was introduced by
him, and a Maxwell's demon. Let us recall the definition of the mutual information between two
systems, $A$ and $B$, in classical information theory:
\begin{equation}\label{def_mutinfo}
I(A:B) = H(A)+H(B)-H(A,B).
\end{equation}
To clarify the quantumness later, we substitute the definition of the joint entropy
$H(A,B)=H(A)+H(B|A)=H(B)+H(A|B)$ into Eq.~(\ref{def_mutinfo}) to define the \textit{locally mutual
information} as
\begin{eqnarray}\label{def_locmutinfo}
J_B(A:B_{\{\ket{B_k}\}}) &=& H(A)+H(B)-H_B(A,B_{\{\ket{B_k}\}}) \nonumber \\
&=& H(A)+H(B)-[H(B)+H(A|B)]_{\{\ket{B_k}\}},
\end{eqnarray}
where the subscript $B$ and $\{\ket{B_k}\}$ are used to stress that we are accessing the system
$B$ locally by using the basis $\{\ket{B_k}\}$. Now the discord is defined as
\begin{eqnarray}\label{def_discord}
\delta(A|B_{\{\ket{B_k}\}}) &=& I(A:B) - J_B(A:B_{\{\ket{B_k}\}}) \nonumber \\
&=& H_B(A,B_{\{\ket{B_k}\}}) - H(A,B).
\end{eqnarray}
Here the (basis-independent) joint entropy $H(A,B)$ is given by the von Neumann entropy of the
whole state $\rho^{A\!B}$, i.e. $H(A,B)=S(\rho^{A\!B})=-\tr \rho^{A\!B}\log_2 \rho^{A\!B}$.

The discord defined here is the work deficit $\Delta$ in Eq.~(\ref{deficit}) when the measurement
is done in the basis $\{\ket{B_k}\}$ only on the subsystem $B$ after one-way communication (from
$A$ to $B$). Zurek described this scenario as the comparison of work-extraction efficiency by
classical and quantum Maxwell's demons: a classical demon is local, while a quantum demon can
perform measurements on the whole system in a global basis in the combined Hilbert space. The
difference in efficiency of work extraction is equal to the discord $\delta(A|B_{\{\ket{B_k}\}})$,
if the classical demon employs $\{\ket{B_k}\}$ as his measurement basis. Thus, the least discord
over all (local) measurement bases, i.e. $\delta(A|B)=\mr{min}_{\{\ket{B_k}\}}
\delta(A|B_{\{\ket{B_k}\}})$, coincides with the work deficit $\Delta$ when only one-way
communication is allowed. Obviously, more communication only helps to increase the classically
extractable work, so $\delta\ge\Delta$ is a general upper bound on $\Delta$.

\subsection{Thermodynamic separability criterion}\label{sec_thermo_separability}
The degree of correlation, particularly the difference between classical and quantum ones, can
also be characterized by discussing the locally extractable work from a heat bath via a given
state, without comparing it with globally extractable work. This is possible despite the fact that
the \textit{optimal} locally extractable work from a pair is the same for both types of
correlation, as we have seen in Section \ref{sec_deficit}: the difference can manifest if we do
not optimize the work in a single setting for extracting. Thus the inequality obtained here works
as an entanglement witness \cite{terhal00} with locally observable thermodynamic quantities.

Suppose that two parties (or demons), Alice and Bob, choose their measurement basis as
$A_\theta=\{P_\theta,P_\theta^\perp\}$ and $B_{\theta'}=\{P_{\theta'},P_{\theta'}^\perp\}$,
respectively, where $\theta \;(\theta')$ represents the direction of the basis. Alice performs her
measurement with $A_\theta$ on her qubits and sends all results to Bob. Then Bob can extract
$1-H(B_{\theta'}|A_\theta)$ bits of work per pair on his side after compressing the information of
his measurement outcomes, where $H(X|Y)$ is the Shannon entropy of $X$, conditional on the
knowledge of $Y$. Only when the shared system is in a maximally entangled state, such as
$\ket{\Phi^+}=(\ket{00}+\ket{11})/\sqrt{2}$, $H(A_\theta|B_\theta)$ can vanish for all $\theta$.
That is, we can extract more work from entangled pairs than from classically correlated pairs.

\begin{figure}
 \begin{center}
  \includegraphics[scale=0.75]{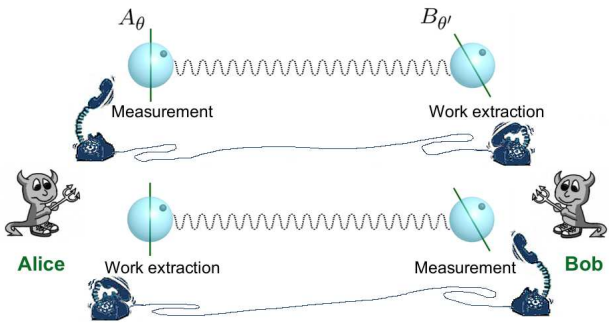}
  \caption{Schematic view of the protocol to extract work from correlated
pairs. Two pairs in the figure represent an ensemble for which Alice and Bob use $A_\theta$ and
$B_{\theta'}$ for their measurement and work extraction. For a half of this ensemble, Alice
measures her state with $A_\theta$ and Bob extracts work from his side along the direction of
$\theta'$, according to Alice's measurement results. For the other half, they exchange their
roles.}
 \label{protocol}
 \end{center}
\end{figure}

Let us choose more measurement bases in order to minimize the dependence of the work on the
particular choice of bases. Alice and Bob first divide their shared ensemble into groups of two
pairs to make the process symmetric with respect to each of them. For each group, they both choose
a projection operator randomly and independently out of a set, $\{A_1,\cdots, A_n\}$ for Alice and
$\{B_1,\cdots, B_n\}$ for Bob, just before their measurement. Then, Alice measures one of the two
qubits in a group with the projector she chose and informs Bob of the outcome as well as her basis
choice. Bob performs the same on his qubit of the other pair in the group. As a result of
collective manipulations on the set of those groups for which they chose $A_i$ and $B_j$, they can
extract a maximum of $2-H(A_i|B_j)-H(B_j|A_i)$ bits of work per two pairs (See
Fig.~\ref{protocol}).

Next, we add up all the work that can be obtained by continuously varying the basis over a great
circle on the Bloch sphere, i.e. the circle of maximum possible size on a sphere. This is similar
in approach to the chained Bell's inequalities discussed in Ref.~\cite{braunstein90}. The circle
should be chosen to maximize the sum. Thus, the quantity we consider is
\begin{equation}\label{var_basis}
\Xi(\rho):=\frac{1}{2\pi}\int_0^{2\pi}\xi_\rho(A(\theta), B(\theta))d\theta,
\end{equation}
where $\xi_\rho(A(\theta),B(\theta^\prime)) =
2-H(A(\theta)|B(\theta^\prime))-H(B(\theta^\prime)|A(\theta))$ is the extractable work from two
copies of $\rho$ in the asymptotic limit when Alice and Bob choose $A(\theta)$ and
$B(\theta^\prime)$, and $\theta$ is the angle representing a point on the great circle. Then, we
can show that $\Xi(\rho)$ can be used as a separability criterion: an inequality
\begin{equation}\label{theineq}
\Xi(\rho)\le\Xi(\ket{00})
\end{equation}
is a necessary condition for a two-dimensional bipartite state $\rho$ to be separable, that is,
$\rho=\sum_i p_i \rho_i^A\otimes\rho_i^B$. The state $\ket{00}$ in the right-hand side of
Eq.~(\ref{theineq}) can be any pure product state $\ket{\psi\psi'}$. We obtained the value of
$\Xi(\ket{00})$ numerically as 0.8854 bits. We refer to this inequality (\ref{theineq}) as a
``thermodynamic separability criterion". The proof of this proposition is based on the concavity
of the entropy \cite{maruyama05a}.

The integral in Eq.~(\ref{var_basis}) can be performed over the whole Bloch sphere, instead of the
great circle, to get another separability criterion. Let $\Xi_{BS}$ denote the new integral, then,
Eq.~(\ref{theineq}) becomes $\Xi_{BS}(\rho)\le \Xi_{BS}(\ket{00})$, where $\Xi_{BS}(\ket{00})$ can
be found numerically as $0.5573$. The proposition above about the separability holds for
$\Xi_{BS}(\rho)$ as well. Let us now compute the value of $\Xi_{BS}(\rho_W)$, where
\begin{equation}
\rho_W=p\ket{\Psi^-}\bra{\Psi^-}+\frac{1-p}{4}\cdot I
\end{equation}
is the Werner state \cite{werner89}, to see the extent to which the inequality can be satisfied
when we vary $p$. It is known that the Bell-Clauser-Horne-Shimony-Holt (Bell-CHSH) inequalities
\cite{chsh69} are violated by $\rho_W$ when $p>1/\sqrt{2}=0.7071$. On the other hand, $\rho_W$ is
inseparable if and only if $p>1/3$, according to the Peres-Horodecki criterion
\cite{peres96,horodecki96}. A bit of algebraic calculation leads to
$\Xi_{BS}(\rho_W)=(1-p)\log_2(1-p)+(1+p)\log_2(1+p)$ and this is greater than $\Xi_{BS}(\ket{00})$
when $p>0.6006$. Therefore, the inequality for $\Xi_{BS}$ is stronger than the Bell-CHSH
inequalities when detecting inseparability of the Werner states. This difference, we suspect, is
due to the nonlinearity of the witness function, which is $\Xi$ in this case. Related analyses are
presented independently in Refs.~\cite{giovannetti04,guhne04a,guhne04b} from the point of view of
the entropic uncertainty relations.

\section{Physical implementations of the demon}
Apart from our own (limited) interests, there are of course a myriad of other interesting works on
Maxwell's demon in the quantum regime, particularly on the physical implementations of the
work-extracting engine under control of the demon. Let us briefly review some of them in this
section.

Lloyd \cite{lloyd97} proposed an experimental realization of Maxwell's demon using nuclear
magnetic resonance (NMR) techniques. A spin-1/2 particle prepared in a standard state, e.g.
$\ket{\!\downarrow}$, works as the demon (memory) to store the information in a given state, which
is also a spin-1/2 particle. Extracting work is done by applying a $\pi$ pulse (to flip it in the
$\{\ket{\!\uparrow},\ket{\!\downarrow}\}$ basis) at the spin's precession frequency $\omega=2\mu
B/\hbar$, where $\mu$ is the magnetic moment of the spin and $B$ is the external magnetic field: a
photon of energy $\hbar\omega$ will be emitted to the field when the spin flips from the higher
energy state $\ket{\!\uparrow}$ to the lower energy state $\ket{\!\downarrow}$. If each of two
spins uses a heat bath at different temperatures, then we can consider a cycle that performs work
in analogy with the Carnot cycle. The quantumness comes into the discussion of the inefficiency of
the cycle, compared with the ideal Carnot case, which is due to the entropy increase by (quantum)
projective measurements. That is, more entropy increase is needed to erase the original
information stored in the demon's memory.

A practical realization of Lloyd's analysis was proposed in Ref.~\cite{quan06a}, in which
superconducting qubits \cite{you05}, instead of spin-1/2 particles, are used to manipulate
information/energy transfer. The necessity of two heat baths at different temperatures is
represented by a temperature gradient between two qubits and the sequence of actual operations is
presented there.

The idea of directly using energy stored in two-level atoms was presented by Scully and his
coworkers \cite{scully01,scully05a,scully05b,scully05c}. In their scenario, two-level atoms are
first randomized in a hohlraum, which is a hollow cavity that thermalizes the energy levels of
incoming atoms, and next they are separated by a Stern-Gerlach-type apparatus into two spatial
paths. Useful energy could be extracted from atoms in the excited state, and then atoms in the two
paths are combined. At the final stage the which-path information is erased isothermally,
dissipating heat (entropy) into the environment, to recycle atoms for another cycle.

In summary, all these physical cycles involve a process that merges two different physical paths
representing two logical states. The merger of two paths corresponds to the logically irreversible
process that we discussed in Section \ref{exorcism}, as was also emphasized in Ref.
\cite{bennett03}, and thus leads to an entropy increase in the outside world of the information
carrier.

Scully \cite{scully03} further proposed another type of heat engine, which has important quantum
aspects. Instead of the two-level atoms in the above example, three-level atoms are now used to
provide useful work (energy) to the radiation field in a cavity. The atom has one excited level
$\ket{a}$ and two nearly degenerate ground levels, $\ket{b}$ and $\ket{c}$. The atoms are
initially prepared to have some small population in the level $\ket{a}$ and a coherent
superposition between $\ket{b}$ and $\ket{c}$, that is, its density operator is given by
$\rho_0=p_a \ket{a}\bra{a} + (1-p_a)\ket{g}\bra{g}$, where $\ket{g}=c_b\ket{b}+c_c\ket{c}$
($|c_b|^2+|c_c|^2=1$). The amplitudes, $c_b$ and $c_c$, as well as the cavity frequency are tuned
so that the probability of transition from $\ket{g}$ to $\ket{a}$ vanishes\footnote{Such a
coherent trapping in $\ket{g}$ occurs due to the destructive quantum interference between two
transitions, namely, $\ket{b}\rightarrow\ket{a}$ and $\ket{c}\rightarrow\ket{a}$. See Chapter 7 of
\cite{quantum_optics_scully}.}.

An interesting consequence of Scully's idea is that quantum coherence, in $\ket{g}$, could be
useful to enhance the efficiency of the thermodynamic cycle even beyond the Carnot efficiency.
This is because it could be possible to extract work, in the form of photons, from a single heat
bath. Such a scenario of extracting work from a single heat bath is indeed reminiscent of
Szilard's demon-assisted one-molecule engine in Section \ref{szilards_engine}. In Scully's engine,
quantum coherence plays the role of the demon. Because of the suppressed absorption of photons by
the atoms, cold atoms absorb less than they would in the absence of coherence, while hot atoms do
emit photons. Hence there is a \textit{sorting action}, which could be seen as a demon's maneuver.
As in Szilard's case, entropy needs to be dissipated when resetting the state of the demon.
Including the entropy cost for initializing the atoms in the total entropy bill ensures the
validity of the second law. A more detailed physical implementation was studied in
Ref.~\cite{quan06b}.

Another noteworthy example, also proposed by Scully, might be the quantum heat engine that makes
use of the difference in energy gaps of a three-level atom \cite{scully02,rostovtsev03}. By
combining maser and laser cavities to control the population of each level, it could be possible
to devise a Carnot-type or Otto-type heat engine and calculate the upper bounds on their
efficiency.

Kieu proposed an idea of a related, but different, type of engine that consists of a two-level
potential well \cite{kieu04}. Work-extracting cycles can be done by manipulating the parameters
for the potential, such as its width and depth. Then the relationships between the temperatures of
the heat baths, the change in energy levels, and the extractable work are analyzed, confirming the
validity of the second law in the quantum regime. This type of idea was considerably extended to
more general cases and scrutinized in terms of quantum Carnot and Otto engines in
Refs.~\cite{quan05,quan07}. The work in Ref.~\cite{quan07} also provides a succinct and
pedagogical presentation of quantum heat engines.

When it comes to the demon in the quantum world, there is also an interesting analysis on
Landauer's erasure principle in the quantum regime. Reference \cite{allahverdyan01} discussed the
validity of the principle when entanglement is taken into account due to the interaction between
the memory system and the environment. However, they identified the Clausius inequality and the
erasure principle directly, and showed that the Clausius inequality could be violated because of
entanglement. This seems to be incompatible with the results in
Refs.~\cite{shizume95,piechocinska00}, where the Clausius inequality was not used to derive the
bound.

\section{Concluding remarks}
Since his birth in the late 19th century, Maxwell's demon has surely been enjoying watching
scientists struggling with his paradox. Nevertheless, he has led us to a new paradigm over the
past century, i.e. the interplay of physics and information theory.

To conclude this article, we wish to re-stress that realizing the irreplaceable reciprocity
between physics and information has given rise to a number of implications in the foundations of
not only quantum mechanics, but also gravity. This may be suggesting that information would help
us merge quantum mechanics and gravity since Maxwell's demon is playing his game at the very core
of both theories. Moreover, the interplay has been a powerful driving force in the development of
quantum information science.

We probably had better prepare for more `demonic' intellectual challenges as more revolutionary
paradigm-shifts might be expected to come in any fields of natural sciences. Therefore it should
be still too early to presume the demise of the demon with plenty of mysteries in nature lying in
front of us.

\section*{Acknowledgments}
This work was supported in part by the National Security Agency (NSA), the Army Research Office
(ARO), the Laboratory for Physical Sciences (LPS), the National Science Foundation (NSF) grant No.
EIA-0130383, the CREST program of the Japan Science and Technology Agency (JST), the Japan Society
for the Promotion of Science Core-to-Core (JSPS-CTC) program, and the JSPS-RFBR grant 06-02-91200.
V.V. thanks the UK Engineering and Physical Sciences Research Council, the Royal Society, and
Wolfson Foundation for support.

\bibliography{the_bib}

\end{document}